\documentclass[useAMS,usenatbib,usegraphicx]{mn2e}
\usepackage{natbib,color,aas_macros}
\usepackage[dvips]{graphicx}
\usepackage{wrapfig}
\graphicspath{{./}}
\usepackage{amsmath}
\usepackage{amssymb}

\def\LaTeX{L\kern-.36em\raise.3ex\hbox{a}\kern-.15em
    T\kern-.1667em\lower.7ex\hbox{E}\kern-.125emX}

\newcommand{\nhcen}{n_{\rm H,cen}}
\newcommand{\msunyr}{{\rm M}_\odot \ {\rm yr}^{-1}}
\newcommand{\msun}{{\rm M}_\odot}
\newcommand{\cc}{{\rm cm^{-3}}}
\newcommand{\HII}{{\rm H_{\, II}}}

\title[Mass of Primordial Stars]
{Primordial star formation under the influence of far ultraviolet radiation: 1540 cosmological halos and the stellar mass distribution}

\author[Hirano et al.]
  {S.~Hirano$^1$\thanks{E-mail: hirano@astron.s.u-tokyo.ac.jp},
  T.~Hosokawa$^2$,
  N.~Yoshida$^{3,4}$,
  K.~Omukai$^5$,
  and 
  H.~W.~Yorke$^6$ \\
  $^1$Department of Astronomy, School of Science, University of Tokyo, Bunkyo, Tokyo 113-0033, Japan\\
  $^2$Department of Physics and Research Center for the Early Universe, University of Tokyo, Bunkyo, Tokyo 113-0033, Japan\\
  $^3$Department of Physics, School of Science, University of Tokyo, Bunkyo, Tokyo 113-0033, Japan\\
  $^4$Kavli Institute for the Physics and Mathematics of the Universe (WPI), Todai Institutes for Advanced Study, \\
  \ \ \ \ the University of Tokyo, Kashiwa, Chiba 277-8583, Japan\\
  $^5$Astronomical Institute, Tohoku University, Sendai, Miyagi 980-8578, Japan\\
  $^6$Jet Propulsion Laboratory, California Institute of Technology, Pasadena CA 91109, USA}

\date{Accepted ?; Received ??; in original form ???}

\pagerange{\pageref{firstpage}--\pageref{lastpage}}
\pubyear{2015}

\begin{document}
\label{firstpage}

\maketitle

\begin{abstract}
We perform a large set of cosmological simulations of early structure formation and follow the formation and evolution of 1540 star-forming gas clouds to derive the mass distribution of primordial stars. 
The star formation in our cosmological simulations is characterized by two distinct populations, the so-called Population III.1 stars and primordial stars formed under the influence of far-ultraviolet (FUV) radiation (Population III.2$_{\rm D}$ stars). 
In this work, we determine the stellar masses by using the dependences on the physical properties of star-forming cloud and/or the external photodissociating intensity from nearby primordial stars, which are derived from the results of 2D radiation hydrodynamic simulations 
of protostellar feedback. 
The characteristic mass of the Pop~III stars is found to be 
a few hundred solar masses at $z \sim 25$, and it
gradually shifts to lower masses with decreasing redshift. 
At high redshifts $z > 20$, about half of the star-forming gas clouds are exposed 
to intense FUV radiation and thus give birth to massive Pop III.2$_{\rm D}$ stars. 
However, the local FUV radiation by nearby Pop III stars becomes {\it weaker} at lower redshifts,
when typical Pop III stars have smaller masses and the mean physical separation 
between the stars becomes large owing to cosmic expansion.
Therefore, at $z < 20$, a large fraction of the primordial gas clouds host Pop III.1 stars. 
At $z \lesssim 15$, the Pop III.1 stars are formed 
in relatively cool gas clouds due to efficient radiative cooling by H$_2$ 
and HD molecules; such stars have masses of a few $\times 10~\msun$. 
Since the stellar evolution and the final fate are determined by the stellar mass, 
Pop~III stars formed at different epochs play different roles in the early Universe.
\end{abstract}


\section{Introduction}
\label{sec:introduction}


Primordial stars play vital roles in the early evolution of the cosmos
by initiating cosmic reionization and chemical enrichment of the inter-galactic medium \citep[e.g.,][]{bromm11}. 
Stellar evolution and death, which regulate the dynamical, radiative, and chemical feedback 
to the surrounding medium, are largely determined by the stellar mass. 
Knowledge of the characteristic mass of primordial stars, or their initial mass function (IMF), 
is thus essential for understanding the initial stages of the first galaxies' formation.
Whereas direct observations can be utilized to derive the stellar IMF in the present-day Universe, 
only limited and indirect observational constraints are available regarding the characteristic mass of the first generation of stars. 
For instance, the elemental abundance patterns of Galactic extremely metal-poor stars provide 
signatures of nucleosynthesis in primordial stars, from which the mass of the progenitor stars can be inferred 
\citep[e.g.,][]{caffau11,aoki14,keller14,tominaga14}.


Theoretical studies offer a possibility to reveal the nature of primordial stars.
The well-established standard cosmological model \citep[e.g.,][]{PLANCK13XVI} allows us to take 
an {\it ab initio} approach to study primordial star formation \citep[see][for recent reviews]{bromm13,glover13}. 
According to these recent studies, primordial star-forming gas clouds are formed 
in the so-called mini-haloes with masses of $10^5 - 10^6 \ \msun$ at very early epochs.
At the centre of such a cosmological mini-halo, 
the gas cloud gravitationally collapses via H$_2$ and/or HD molecular cooling, 
so that a tiny ($\sim 0.01 \ \msun$) embryo -- protostellar core -- forms \citep[e.g.,][]{yoshida08}.


The newly-born protostellar core grows in mass via rapid gas accretion from the surrounding envelope. 
The typical mass of the accretion envelope is very close to that of 
the original gravitationally-unstable gas cloud (so-called Jeans mass), 
\begin{eqnarray}
M_{\rm Jeans} \approx 1000~\msun \left( \frac{T}{200 \ {\rm K}} \right)^{3/2} \left( \frac{n_{\rm H}}{10^4 \ {\rm cm^{-3}}} \right)^{-1/2} \ , 
\label{eq:M_BE}
\end{eqnarray}
where $T$ is the gas temperature and $n_{\rm H}$ is the hydrogen number density \citep{abel02}. 
The final stellar mass critically depends on how long the gas accretion continues. 
A key process in this later accretion stage is the protostellar radiative feedback, 
which sets the final stellar mass by terminating the mass accretion. 
\cite{mckee08} evaluate the potential impacts of the ultraviolet (UV) radiative 
feedback using a semi-analytical model. They consider that, when a protostar becomes 
massive enough to emit a copious amount of ionizing photons, an $\HII$ region grows 
in polar directions of a circumstellar disc. The stellar ionizing photons heat up and 
photoevaporate the gas on the disc surface. 
More recently, the feedback process has been studied with multidimensional 
radiation hydrodynamic (RHD) numerical simulations \citep{hosokawa11,stacy12,susa13}, 
demonstrating that UV radiative feedback does indeed shut off the mass accretion and the 
resulting stellar masses can be a few tens of solar masses.


These previous studies calculate only several individual cases and thus it remains unclear if they represent fiducial cases. Obviously, statistical studies are needed to study the overall mass distribution of primordial stars. 
Our previous work \citep[][hereafter Paper I]{hirano14a} achieves this goal with more than one hundred cosmological samples of primordial star-forming clouds. 
We first follow the early evolution until the formation of the protostellar core with three-dimensional (3D) cosmological simulations. 
We then study the subsequent evolution in the mass accretion stage under the influence of protostellar UV feedback by performing local 2D RHD simulations \citep{hosokawa11}. 
The protostellar evolution is calculated self-consistently by numerically solving the interior structure of the protostar \citep[e.g.,][]{omukai03a}. 
A wide range of stellar masses is obtained, extending from 10 to 1600 $\msun$, which suggests a great diversity of primordial stars.
Despite several physical effects that are not directly realized in 2D simulations, 
e.g., disc fragmentation and stellar multiplicity, such diversity is consistently 
reported by other studies including recent 3D simulations \citep{susa14}.
\footnote{
We adopt the so-called alpha-viscosity whose value is calibrated 
with respect to the results of recent 3D simulations (see appendix A in Paper I). 
Thus our 2D calculations include efficient 
angular momentum transfer via the non-axisymmetric disc structure.}
Interestingly, we find correlations between the final stellar masses
and the physical properties of star-forming clouds, 
\begin{eqnarray}
M_{\rm *} = 100 \ {\rm M_{\odot}} \left( \frac{\dot{M}_{\rm Jeans}}{2.8 \times 10^{-3} \ {\rm M_{\odot} \ yr^{-1}}} \right)^{0.8} \ ,
\label{eq:MIII1_cloud_paper1}
\end{eqnarray}
where $\dot{M}_{\rm Jeans}$ is the gas infall rate at the Jeans scale, 
as well as between the final stellar masses and the physical properties of the mini-haloes, 
\begin{eqnarray}
M_{\rm *} = 100 \ {\rm M_{\odot}} \left( \frac{1+z}{20} \right)^{3} \left( \frac{M_{\rm vir}}{{\rm 3 \times 10^5 \ M_{\odot}}} \right)^{2} \ ,
\label{eq:MIII1_virial_paper1}
\end{eqnarray}
where $z$ is the forming redshift and $M_{\rm vir}$ is the virial mass (these equations are given in eqs.~13 and 19 in Paper I). 
Such correlations are useful 
because one can estimate the final stellar mass using the early physical state of the clouds or of the haloes, 
without following the detailed evolution from the cloud collapse up to the termination of mass accretion on to the protostellar core.


The ultimate goal of our statistical study is to directly calculate the primordial IMF theoretically. 
Clearly, a number of improvements over previous studies are needed to achieve this. 
First, in Paper I, we select star-forming gas clouds from a few different simulations (realizations) with small cosmological volumes. 
It would be desirable to use a single large volume to make a cosmologically representative sample.
Secondly, it is often assumed that the primordial stars are unaffected by any external feedback from other nearby stars 
(Population III.1 stars~\footnote{In Paper I, we have classified Pop III.1 stars into two sub-classes 
depending on whether HD molecular cooling affects the thermal evolution of the cloud 
(Pop III.1$_{\rm HD}$) or not (Pop III.1$_{\rm H_2}$).}). 
However, primordial stars forming under the influence of a UV radiation background, the so-called Population III.2 stars, 
are thought to have different characteristics from Pop III.1 stars,
because the abundance of H$_2$ and HD molecules, which determine the thermal evolution during the cloud collapse, 
are vulnerable to the external radiation. 
The relative occurrence of the two populations is not known.

In this paper, we consider a comoving volume of (3 $h^{-1}$~Mpc)$^3$ for our 
cosmological simulations, rather than the simulation volume of Paper I.
Furthermore, we classify the Pop III.2 stars into two subclasses depending on the hardness of the radiation: 
(1)~the photodissociation-dominated case (Pop III.2$_{\rm D}$) and 
(2)~the photoionization-dominated case (Pop III.$2_{\rm I}$), 
where the subscripts ${\rm D}$ and ${\rm I}$ stand for dissociation and ionization, respectively. 
In the former case, because H$_2$ and HD molecules are destroyed by photodissociating photons, 
the temperature in a collapsing cloud is higher than for the Pop III.1 case 
\citep[e.g.,][]{omukai01b,omukai03b,o'shea08}. 
The mass accretion rate on to the protostar increases
following the well-known scaling relation $\dot{M} \propto T^{3/2}$. 
At the higher accretion rates, the protostar has a larger
radius and lower effective temperature for a given stellar mass \citep[the ``bloating'' phenomenon, see e.g.][]{zinnecker07,hosokawa11}.
This weakens the stellar UV feedback, and the resulting final
stellar mass becomes higher than in the Pop III.1 case.


In the latter case (Pop III.$2_{\rm I}$), radiation fields affect the thermal evolution in the opposite direction; 
photoionization first enhances the abundance of free electrons, 
which are the catalyst for generating H$_2$ molecules, 
and then a collapsing cloud evolves at low temperature 
due to the enhanced H$_2$ and HD formation and cooling \citep[e.g.,][]{yoshida07,hosokawa12b}. 
Overall, these theoretical studies suggest that 
the mass of a primordial star depends critically on the local strength of the radiation emitted by other stars.

Clearly, we need to consider all sub-populations in order to calculate the primordial IMF. 
In the present study, we consider both Pop III.1 and III.2$_{\rm D}$ stars 
but ignore the contribution of Pop III.$2_{\rm I}$ stars (see Sec. \ref{sec:discussion} for a discussion of this topic). 
In order to generate non-biased cosmological samples of star-forming clouds, 
we first choose 1540 haloes in our large cosmological simulation. 
We then derive the correlations between the stellar mass and the
physical properties of the parent gas cloud to estimate the mass of a star forming in each cloud. 
To this end, we investigate how the final stellar mass depends on the
strength of the photodissociating radiation field
by performing computationally intensive RHD simulations for 45 representative cases. 
We then apply the derived formulae for the final stellar mass to our cosmological samples.
For each cloud, the local intensity of the photodissociating radiation is calculated 
by summing up contributions from nearby primordial stars distributed in the cosmological volume. 
This approach provides us with mass distributions of Pop III.1 and III.2$_{\rm D}$ stars at different epochs.


The remainder of the paper is organized as follows. 
In Section \ref{sec:method}, we describe the numerical methods employed in our set of simulations: 
cosmological simulations, RHD simulations, and the post-processing calculations of the local far-ultraviolet (FUV) intensity. 
In Section \ref{sec:overview}, we first summarize our main results. 
Section \ref{sec:sampling} shows the statistics of our cosmological samples of the primordial star-forming regions. 
Section \ref{sec:III2D} presents the results of the local RHD simulations, 
with which we also explain how the stellar mass is assigned to each primordial cloud. 
Section \ref{sec:distribution} shows the resulting mass distributions of primordial stars, 
classifying our statistical samples into Pop III.1 and III.2$_{\rm D}$ cases. 
We finally give concluding remarks in Section \ref{sec:discussion}.


Throughout this paper, we adopt the standard 
$\Lambda$-cold dark matter ($\Lambda$-CDM) cosmology 
with the total matter density $\Omega_{\rm m} = 0.3086$, the baryonic density $\Omega_{\rm b} = 0.04825$, the dark energy density $\Omega_{\Lambda} = 0.6914$ in units of the critical density, the Hubble constant $h = 0.6777$, and the primordial index $n_{\rm s} = 0.9611$ \citep{PLANCK13XVI}. The power spectra are defined by the equation in \cite{eisenstein99} and normalized to $\sigma_{8} = 0.8288$.

\section{Methods}
\label{sec:method}

\begin{figure*}
\begin{center}
\resizebox{16.5cm}{!}{\includegraphics[clip,scale=1]{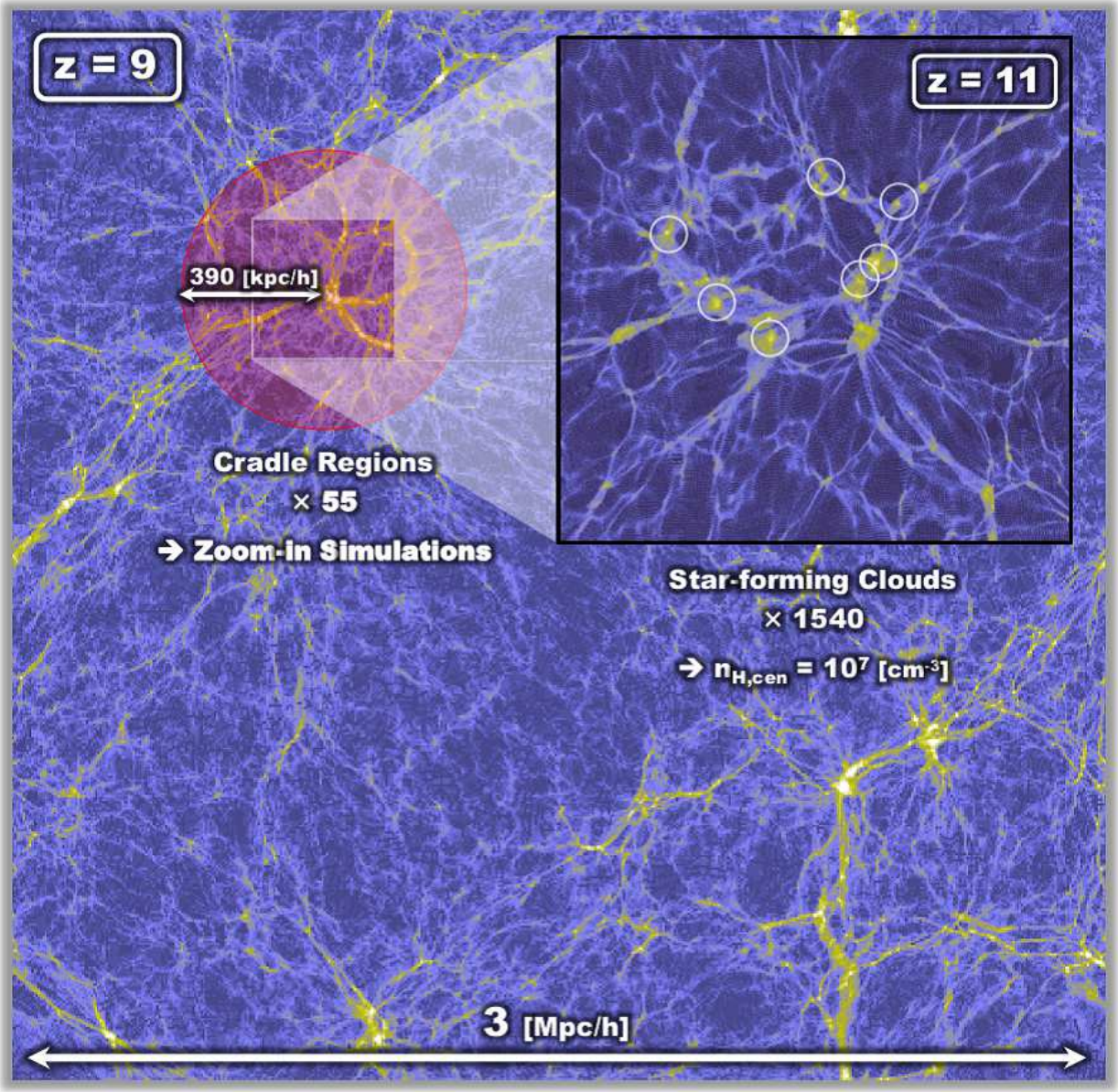}}
\caption{
Schematic view of the cosmological simulation. 
From a parent simulation (the background panel), 
we generate 55 zoomed initial conditions which cover each cradle region 
(the red circle shows one of them). 
We finally obtain 1540 star-forming clouds from zoom-in simulations 
(white circles in the foreground panel)
and follow their formation and evolution. 
The colour contour reveals the projected density level of DM 
which increase from blue to yellow.}
\label{fig:plot_pre-zoom}
\end{center}
\end{figure*}


\subsection{Sampling Star-Forming Regions in Cosmological Simulations}


First, we perform a large parent cosmological simulation with relatively low spatial resolution and identify dark matter (DM) haloes. 
We then recalculate the evolution using the so-called hierarchical zoom-in technique for 55 cradle regions 
which contain a lot of haloes. 
From these zoom-in simulations, we pick out the star-forming clouds and
calculate the subsequent collapse until the central density reaches $10^7 \ {\rm cm^{-3}}$. 
The cloud state at this moment is used to infer the final stellar mass (see Sec. \ref{ssec:recst_formula}). 
Here, we describe each step of our procedure, emphasizing
newly developed techniques and the improvements over previous studies.


Cosmological simulations are performed using
the parallel $N$-body / smoothed particle hydrodynamics (SPH) solver {\sc gadget-2} \citep{springel05} 
in its version suitably adopted for primordial star formation. 
We solve chemical rate equations for 14 primordial species 
(${\rm e^-}$, ${\rm H}$, ${\rm H^+}$, ${\rm H^-}$, ${\rm He}$, ${\rm He^+}$, ${\rm He^{++}}$, 
H$_2$, ${\rm H_2^+}$, ${\rm D}$, ${\rm D^+}$, ${\rm HD}$, ${\rm HD^+}$, ${\rm HD^-}$) as in \cite{yoshida06, yoshida07}. 
We have updated the cooling rates for H$_2$ and HD \citep[see][]{galli13} as well as the three-body H$_2$ formation rates \citep{forrey13a, forrey13b}.

\begin{figure}
\begin{center}
\resizebox{7.5cm}{!}{\includegraphics[clip,scale=1]{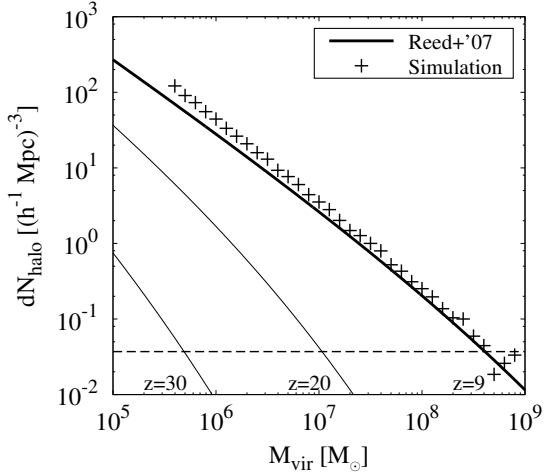}}
\caption{
Mass function of DM haloes. 
The crosses show the result of the parent simulation at $z = 9$. 
We run the FOF halo finder with linking length $b = 0.2$. 
We retain haloes containing more than 50 $N$-body particles. 
The solid lines show the analytically calculated Press-Schechter mass function at $z = 9$ (thick), 20, and 30 (thin) using the \citet{reed07} code. 
The dashed horizontal line represents the number density of $1~[(3~h^{-1}~{\rm Mpc})^{-3}]$, with which one halo should be found in the cosmological volume of the parent simulation.
}
\label{fig:plot_HMF-KPC_M0768_L3MPC}
\end{center}
\end{figure}

We generate the cosmological initial conditions using the modified version of {\sc n-genic} \citep{springel05}. 
In order to achieve sufficient spatial resolution to resolve the
primordial gas clouds with masses of $\sim$ 1000 $\msun$, we use a hierarchical zoom-in technique. 
First, we run the parent $N$-body (DM-only) simulation with a large cosmological volume of $L_{\rm box} = 3 \ h^{-1}$ Mpc (comoving)~\footnote{Other length scales described below are all in comoving scale.} 
on a side employing $768^3$ DM particles. 
The particle mass of the dark matter component is 7400 $\msun$ and the
gravitational softening length is set to be 115 pc. 
We follow the evolution from $z = 99$ to 9. 
The background panel in Fig. \ref{fig:plot_pre-zoom} shows the final output of this simulation. 
Primordial clouds form within the dark matter haloes located at the densest parts of the cosmic filaments. 
We then pick out all candidate sites for primordial star formation in the simulated volume. 
We identify DM haloes by running the Friends-Of-Friends (FOF) halo finder. 
Fig. \ref{fig:plot_HMF-KPC_M0768_L3MPC} shows the obtained mass function of DM haloes. 
Our simulation result agrees well with the analytical prediction (Press-Schechter mass function), 
confirming that our parent simulation represents a typical cosmological volume. 
Note that our halo selection should be non-biased if all of the star-forming sites are selected from this simulation volume.


Fig. \ref{fig:plot_pre-zoom} shows that the dark haloes are strongly clustered. 
We locate 55 zoom-in spherical regions with the radius of 390 $h^{-1}$ kpc 
to cover such cradle regions 
(one of them is marked with the red circle in Fig. \ref{fig:plot_pre-zoom}). 
We only use the inner 3/4 volume of each zoom-in region to look for gas clouds 
in order to exclude non-physical haloes which appear in the outer part of the zoom-in regions.
The net volume considered as primordial star-forming regions is about 20 per cent of the total simulation volume. 
We generate 55 zoomed initial conditions with baryonic components 
(SPH particles) with increased mass resolution 
by a factor of $4^3$.
The resulting effective number of particles in a zoom-in region is $3072^3$. 
The corresponding small-scale density fluctuations are added suitably 
as given by our adopted $\Lambda$-CDM cosmology. 
The masses of DM particles and of the gas particles 
within the zoom-in region are 116 and 19~M$_\odot$, respectively.
The gravitational softening length is set to be 29 pc. 
We continue the zoom-in simulations down to redshifts $z \simeq 30 - 25$, 
when the first gravitationally collapsing cloud is formed.


When non-linear objects are formed, we further clip the densest part out of the zoom-in regions with the following cut-out criteria: 
\begin{itemize}
\item[1.] 
Because we stop each zoom-in simulation when a cloud begins to collapse, the subsequent evolution of the other clouds is not followed in the same single simulation.
We tag potential sites where other clouds are later formed, by identifying the regions with densities a few times greater than the cosmic mean value. 
These zoomed-in regions normally have $\sim 10 - 100$ candidates.

\item[2.] 
We select haloes which lie in the densest region within 1 physical kpc radius around the density peak,
implicitly assuming that the halo forming there will not be affected by
external feedback from nearby sources except the photodissociating feedback,
including photoionizing, chemical, and dynamical feedbacks.
Primordial clouds forming at the centres of such haloes will host Pop III.1 or III.2$_{\rm D}$ stars.
\end{itemize}

Finally, we restart the simulations for each selected gas cloud. 
When a cloud collapses, we continually increase the spatial resolution by adopting 
the particle-splitting technique \citep{kitsionas02} to insure
that the local Jeans length is always resolved 
by 15 times the local smoothing length of the SPH particles.
We follow the collapse of each cloud until its central density reaches $10^7 \ {\rm cm^{-3}}$,
which ultimately results in 1540 primordial star-forming regions within one parent cosmological volume.


In the present study, we do not consider the gas metal enrichment and
hence the formation of Pop II stars.
Previous studies show that the FUV radiation from nearby Pop II stars 
dominates over the Pop III component at $z \lesssim 10 - 15$ \citep[e.g.,][]{agarwal12,johnson13}. 
Thus, we stop our simulations at $z = 10$ 
(see also discussion in Sec. \ref{ssec:uncertainties}).

\subsection{Deriving the Correlations between the FUV Intensity and Stellar Masses: 2D RHD Simulations}
\label{ssec:rhdsimu}

The mass of a Pop III.1 star is largely determined by the physical properties of the natal gas clouds, 
e.g., the cloud mass and the rotation degree as shown in Paper I. 
For Pop III.2$_{\rm D}$ stars, however, there are additional parameters 
that could affect the star formation process. 
One is the intensity of photodissociating FUV radiation in Lyman-Werner (LW) bands, which is often normalized 
as $J_{21} = J_{\rm LW} / (10^{21} \ {\rm erg \ s^{-1} \ Hz^{-1} \ sr^{-1}})$. 
This radiation destroys H$_2$ and HD molecules, and prevents the cooling and collapse of the cloud. 
In this study, we investigate the dependence of Pop III.2$_{\rm D}$ star-formation on the parameter $J_{21}$.  

Another parameter is the central density when the photodissociating photons reach the cloud. 
We fix this parameter at $n_{\rm H,cen,rad} = 10 \ {\rm cm^{-3}}$ for the following reasons.
In reality, the FUV photons might affect the earlier phase of the collapse ($n_{\rm H,cen,rad} < 10 \ {\rm cm^{-3}}$).
Pop III.2$_{\rm D}$ stars could then form with less efficient self-shielding. 
However, in our calculations it takes $\sim$100~Myr 
for such a low-density cloud to begin to collapse with the reduced amount of H$_2$ molecules. 
It is unlikely that the intensity of FUV radiation remains strong enough for the Pop III.2$_{\rm D}$ star formation mode for such long time; typical stellar lifetimes of the FUV sources are only a few million years. 
For the opposite case ($n_{\rm H,cen,rad} > 10 \ {\rm cm^{-3}}$), 
self-shielding prevents photodissociation and the cloud forms a Pop III.1 star;
we find that self-shielding becomes sufficiently efficient in the cloud at densities $\sim 10^2 - 10^3 \ \cc$. 
We thus adopt the critical value $n_{\rm H,cen,rad} = 10 \ \cc$.


We calculate the evolution of nine different clouds irradiated by
photodissociating photons. We study the effects of external irradiation,
varying the intensity as a free parameter with
$J_{21}$ = 0, 0.1, 0.316, 1, and 10.
We have explicitly checked that for $J_{21}$ = 0.01, 
the thermal evolution of the collapsing cloud is nearly identical to the Pop III.1 case ($J_{21}$ = 0). 
On the other hand, the collapse is almost completely prohibited for intensities exceeding $J_{21} = 10$. 
We thus examine the parameter range of $J_{21} = 0.1 - 10$.


We study the formation of Pop III.2$_{\rm D}$ stars in a two-step manner as follows. 
First, we calculate the evolution of collapsing clouds under the FUV fields by {\sc gadget-2}. 
A uniform FUV radiation field is assumed here. 
We employ the self-shielding functions \citep{wolcott-green11a,wolcott-green11b} for H$_2$ and HD molecules. 
For each SPH particle, we calculate the column densities along six directions ($\pm X,\ \pm Y,\ \pm Z$) 
to account for the directional dependence of the self-shielding effect. 
We stop the collapse simulations when the central density reaches $n_{\rm H,cen} = 10^{13} \ {\rm cm^{-3}}$
and assume that an optically thick hydrostatic object (protostar) has formed at this point. 
The final configuration is used to generate
the {\it initial} conditions for our 2D RHD grid-based simulations of the subsequent protostellar accretion stage
by azimuthal averaging around the rotational axis ($\phi$-direction).


We follow the subsequent evolution after the birth of the protostar by
conducting 2D axisymmetric RHD simulations of the collapsing cloud
coupled with the simultaneous stellar evolution of the central accreting protostar located in a central sink cell \citep[][Paper I]{hosokawa11,hosokawa12b}. 
We follow the evolution until the mass accretion rate falls 
below $10^{-4} \ \msunyr$ in each case. 
We regard the mass at this moment as the final stellar mass. 
Whereas we normally follow the accreting protostar's evolution by
solving for the stellar interior structure, it occasionally becomes difficult
to construct a stellar model with highly variable mass accretion histories.
This often happens in Pop III.2$_{\rm D}$ cases, when the mass accretion
rates are relatively high (see Sec. \ref{sec:III2D}) and the protostar
has an extremely large radius \citep[e.g.,][]{hosokawa12a}.
In order to continue the calculations in these cases, we switch to an analytic stellar model, which
agrees with the numerical results well, only when the accretion rate exceeds 
$10^{-2}~\msunyr$ (see also Appendix \ref{app:spgp}).

\begin{figure*}
\begin{center}
\resizebox{15cm}{!}{\begin{tabular}{cc}
\includegraphics[clip,scale=1]{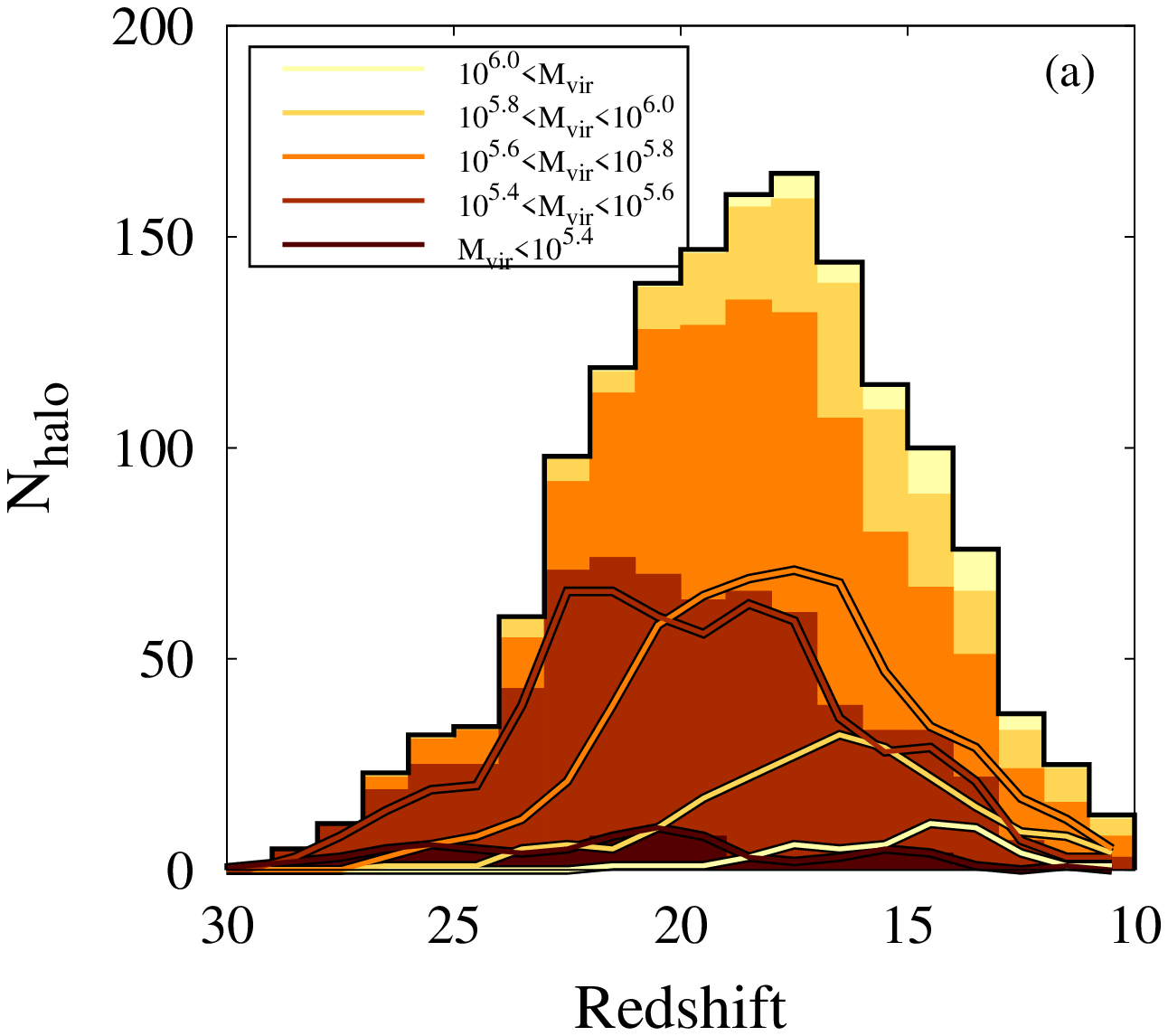}&
\hspace{-15mm}
\includegraphics[clip,scale=1]{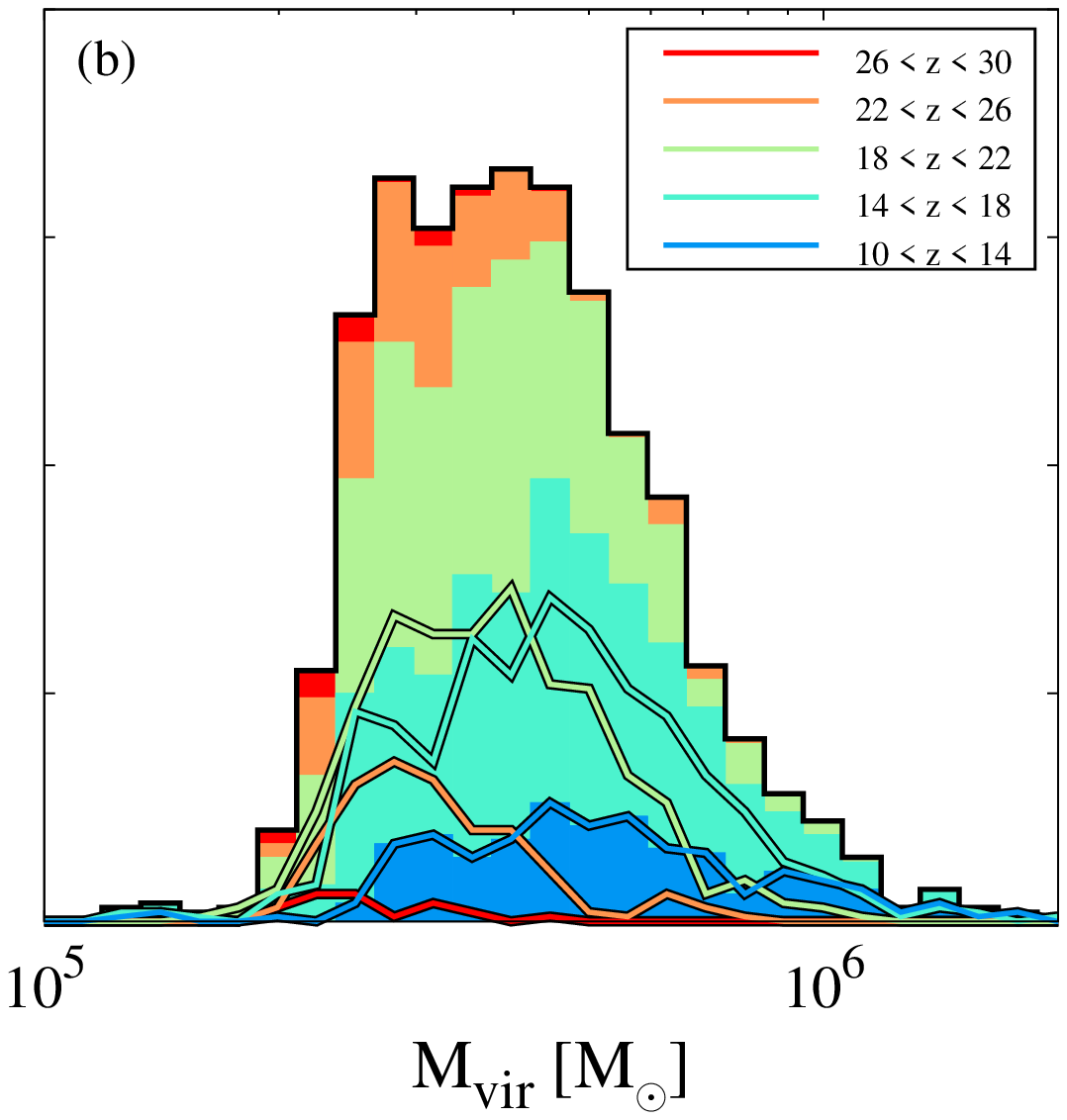}
\end{tabular}}
\caption{
Distributions of DM mini-haloes that host primordial star-forming clouds as functions of the formation redshift (left-hand panel) or the virial mass (right). 
In the left-hand panel, the bin width is $\Delta z = 1$ and the different colours indicate the different ranges of the virial masses: 
$M_{\rm vir}/{\rm M}_\odot<10^{5.4}$, 
$10^{5.4}<M_{\rm vir}/{\rm M}_\odot<10^{5.6}$, 
$10^{5.6}<M_{\rm vir}/{\rm M}_\odot<10^{5.8}$, 
$10^{5.8}<M_{\rm vir}/{\rm M}_\odot<10^{6.0}$, and 
$10^{6.0}<M_{\rm vir}/{\rm M}_\odot$. 
In the right-hand panel, the bin width is logarithmically equal, $\Delta M_{\rm vir}(M) = (10^{0.1} - 1) \times M$, and the different colours indicate the different ranges of the redshifts:
$10<z<14$, $14<z<18$, $18<z<22$, $22<z<26$, and $26<z<30$.
The black lines show the sums over all virial masses or redshifts.
The redshifts and virial masses are measured when $\nhcen = 10^7~\cc$.
}
\label{fig:plot_z-Mvir}
\end{center}
\end{figure*}

\subsection{Classification of Star Formation Modes: Evaluation of Local ${\bf J_{LW}}$}
\label{ssec:j21calc}


To derive the primordial stellar mass distribution including both 
Pop III.1 and III.2$_{\rm D}$ stars, 
we need to evaluate the local intensity of the photodissociating
radiation at each cloud in a self-consistent manner 
with the spatial distribution of the primordial stars. 
For this purpose, we perform the following post-processing calculations:


\begin{itemize}
\item[1.] 
The primordial gas clouds are sorted by their formation redshifts $z_{*1}$, the epoch when the cloud's central density reaches $10^7 \ \cc$.
We first begin with assigning the stellar mass to the halo which has the highest $z_{*1}$.

\item[2.] 
We then evaluate the local FUV intensity for the clouds forming at the lower redshifts.
We assume that the star radiates at a constant luminosity $Q_{\rm LW}$ during its lifetime $t_*$ and dies at redshift $z_{*2} = z_{*1} - \Delta z(t_{*})$ and adopt the numerical results of \citet{schaerer02}, who present the values of $Q_{\rm LW}$ and $t_*$ as functions of the stellar mass.
We evaluate the FUV intensity when a cloud is in the pre-collapse stage $z_{\rm ini}$ ($n_{\rm H,cen} = 10 \ \cc$). 
We count the FUV intensity from stars which have $z_{*1} > z_{\rm ini} > z_{*2}$ to obtain the local field.

\item[3.] 
The stellar mass in a halo forming at the lower redshift is determined using the correlation derived from the 2D RHD simulations.
We classify the star as Pop III.2$_{\rm D}$, if the local FUV intensity is higher than the critical value $J_{\rm 21,crit} = 0.1$. 
The mass of a star formed within the irradiated cloud is 
largely determined by the cloud's properties for Pop III.1 
cases, whereas it is also dictated by the local FUV intensity for Pop III.2$_{\rm D}$ cases.

\item[4.] 
We repeat steps 2 and 3 for the clouds one by one in the order of decreasing formation redshift.
The stellar masses are accordingly determined by the above procedure.
\end{itemize}

\begin{figure}
\begin{center}
\resizebox{7.5cm}{!}{\includegraphics[clip,scale=1]{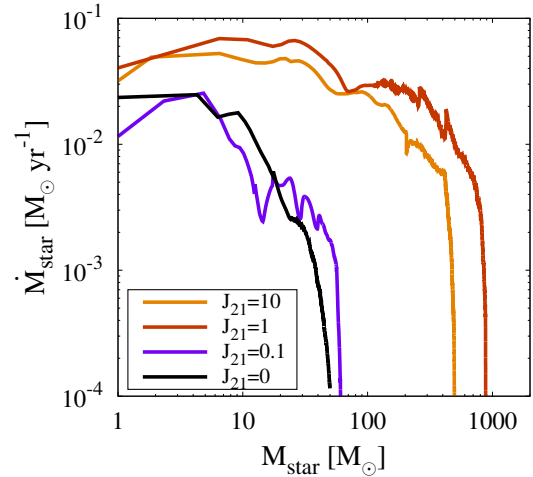}}
\caption{
Accretion histories of mass accreting protostars, 
which form from the same cloud (ID = 4 in Table \ref{tab:M_III2DIS}) but
with different FUV intensities $J_{21}$ = 0, 0.1, 1, and 10, 
as a function of the stellar mass. 
When the mass accretion rates fall below $10^{-4}$~M$_\odot \ {\rm yr^{-1}}$, 
we stop the simulations and assume the masses at this moment as the final 
stellar masses (50, 61, 883, and 496~M$_\odot$, respectively).}
\label{fig:plot_III2-EXTEL_Matar-dMdt_HFS032}
\end{center}
\end{figure}

\begin{figure*}
\begin{center}
\resizebox{18cm}{!}{\includegraphics[clip,scale=1]{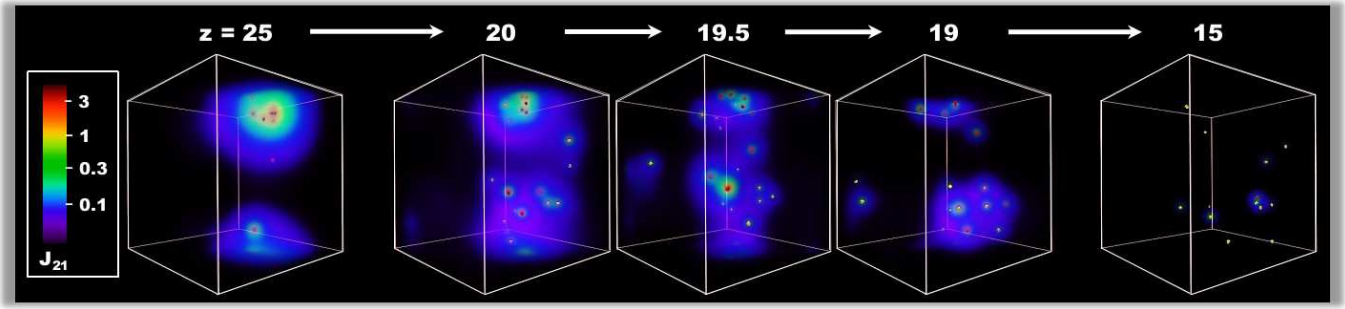}}
\caption{
The local FUV intensity field, $J_{21}$, in the same comoving cosmological volume with $(3~h^{-1}~{\rm Mpc})^3$ at $z = 25$, 20, 19.5, 19, and 15. 
The colour contours indicate the FUV intensity ranging from $J_{21} = 0.025-6.3$ (blue to red). 
The yellow and red clumps show the active Population III.1 and III.2$_{\rm D}$ stars, respectively.
The local FUV radiation field decreases with decreasing redshift (from left- to right-hand panels), because the typical stellar mass becomes lower and separations between the stars are stretched by the cosmic expansion.
}
\label{fig:J21MAP}
\end{center}
\end{figure*}


The intensity of the FUV radiation at the location of the $i$th cloud 
contributed by the star in the $j$th cloud is
\begin{eqnarray}
J_{{\rm LW},i}^j = \frac{f_{\rm esc}}{\pi} \frac{h \nu_{\rm avg}}{\Delta \nu_{\rm LW}} \frac{Q_{{\rm LW},j}}{4 \pi d_{ij}^2} \ , 
\label{eq:J_LW_local_j-th}
\end{eqnarray}
where 
$f_{\rm esc}$ is the fraction of LW photons that escape from the $j$th halo, 
$h \nu_{\rm avg}$ is the average photon energy emitted from a Pop III star in the LW band, 
$\Delta \nu_{\rm LW}$ is the difference between the maximum and minimum frequencies of 
the LW bands,\footnote{$\nu_{\rm avg} = \nu_{\rm max} =$ 13.6 eV ($3.288 \times 10^{15}$ Hz) and $\nu_{\rm min} =$ 11.2 eV ($2.708 \times 10^{15}$ Hz).} and 
$d_{ij}$ is the distance between the centres of the $i$th and $j$th clouds \citep{agarwal12}. 
According to \cite{kitayama04}, the escape fraction $f_{\rm esc}$ should be around $0.1 - 1$ 
for the mini-haloes considered with $M_{\rm vir} = 2 \times 10^5 - 2 \times 10^6$ M$_\odot$. 
We adopt the fiducial value of $f_{\rm esc} = 0.5$. 
The local FUV intensity at the $i$th cloud $J_{21,i}$ is then calculated by summing up the contributions 
from all the nearby clouds with $z_{*1,j} > z_{{\rm ini},i} > z_{*2,j}$,
\begin{eqnarray}
J_{{\rm LW},i} = \sum_j^N J_{{\rm LW},i}^j \ .
\label{eq:J_LW_local_total}
\end{eqnarray}
We only consider the contemporary sources in the simulation box and 
ignore the background radiation (see Sec. \ref{sec:distribution_ssec:classification}).

\section{Overview}
\label{sec:overview}


We present an overview of our main results in this section. 
Fig. \ref{fig:plot_z-Mvir}(a) shows the redshift distribution of 1540 mini-haloes located in our cosmological simulation.
Our sample ranges from $z = 30$ to 10 and peaks around $z \sim 20$. 
Fig. \ref{fig:plot_z-Mvir}(b) shows the mass distribution of the haloes,
which indicates a peak around $M_{\rm vir} \simeq 3 \times 10^5 \ {\rm M}_\odot$. 
The samples in Paper I did not show such a peak, indicating some selection bias associated with the small cosmological volume and 
the limited statistics due to the small number of samples. 
In Paper I, we show that the halo mass is the key parameter to determine the stellar mass. 
Thus, the more accurate halo mass distribution of the current samples can be expected to yield a modified stellar mass distribution.


\begin{figure}
\begin{center}
\resizebox{7.5cm}{!}{\includegraphics[clip,scale=1]{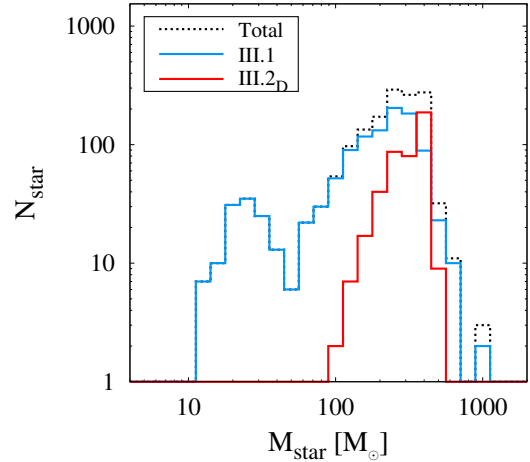}}
\caption{
Mass distribution of Pop III.1 and III.2$_{\rm D}$ stars. 
The dotted line shows the sum of the two populations
(see also Fig. \ref{fig:plot_IMF_dz} below).
}
\label{fig:plot_IMF}
\end{center}
\end{figure}

Fig. \ref{fig:plot_III2-EXTEL_Matar-dMdt_HFS032} shows 
the mass accretion histories on to a protostar, obtained from our 2D RHD simulations. 
As described in the previous sections, the computationally intensive RHD calculations are needed to examine 
the correlation between the final stellar mass and the intensity of the photodissociating radiation. 
In all of the cases presented, the mass accretion rate decreases sharply at some point: 
i.e., when the protostellar feedback shuts off mass accretion. 
In general, for the larger $J_{21}$, the gas temperature 
in the accretion envelope is higher \citep[e.g.,][]{omukai01a,o'shea08}, 
causing higher accretion rates on to the protostar. 
Indeed, we find that the final stellar mass is larger for higher FUV intensities. 
We will account for this effect when estimating the final stellar masses for a given FUV intensity 
in order to construct the mass distribution of Pop III.2 stars (see Sec. \ref{sec:III2D}).


Finally, we classify our data of 1540 primordial clouds into Pop III.1 and III.2$_{\rm D}$ cases, 
according to our criteria based on the local photodissociating FUV intensity ($J_{21} < 0.1$ or not). 
Fig. \ref{fig:J21MAP} shows 
the spatial distributions of the FUV intensity $J_{21}$ and the locations of 
active Pop III.1 (yellow dots) and III.2$_{\rm D}$ stars (red dots) in the cosmological volume. 
There are small clusters of Pop III.2$_{\rm D}$ stars at high redshifts, 
where the photodissociating radiation is very strong. 
We determine the stellar masses from the cloud properties and the computed FUV intensity. 
Fig. \ref{fig:plot_IMF} shows the mass distributions of primordial stars including Pop III.1 and III.2$_{\rm D}$ stars. 
The Pop III.1 mass distribution has two peaks at $\simeq 250$ and $25 \ \msun$. 
The two characteristic masses reflect different paths of thermal evolution during the early runaway collapse stage, 
which are driven by either H$_2$ cooling alone or H$_2$ cooling plus HD cooling (Paper I). 
Pop III.2$_{\rm D}$ stars have relatively large masses, clustering around $\simeq 400 \ \msun$. 
We discuss the redshift dependence of the mass distribution in Section \ref{sec:distribution}.

\begin{figure}
\begin{center}
\resizebox{7.5cm}{!}{\includegraphics[clip,scale=1]{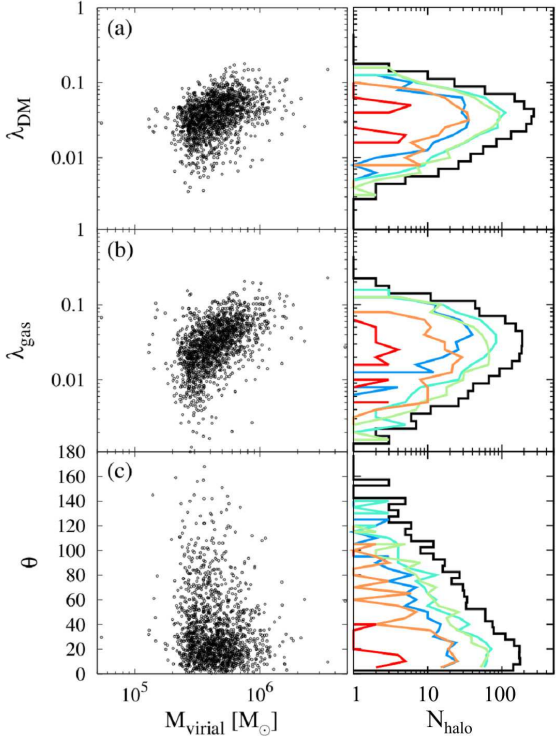}}
\caption{
Scatter plots and histograms of basic properties of the gas clouds on the virial scale. 
The left-hand panels show, from top to bottom, spin parameters of DM, spin parameters of baryonic components, and 
the offset angles between angular momentum vectors of the two components. 
The right-hand panels show their histograms; the colours represent the same redshift ranges 
as in the right-hand panel of Fig. \ref{fig:plot_z-Mvir}.}
\label{fig:plot_2KPC_spin-left+right_half}
\end{center}
\end{figure}

\section{Cosmological Sample of Primordial Star-Forming Clouds}
\label{sec:sampling}


We study the physical properties of the 1540 primordial star-forming clouds found in our cosmological simulation. 
In this section, we ignore the effect of FUV fields and assume that all the clouds bear Pop III.1 stars, allowing
a direct comparison with the statistical results presented in Paper I. 
The dependences of cloud properties and stellar mass distribution on the formation redshifts are discussed.


The cloud properties are calculated at two different mass scales by averaging 
over the gas in the virialized DM haloes and 
over the gravitationally collapsing gas at the Jeans scale. 
We define the boundary for the former to be the halo virial radius, 
within which the average matter density is 200 times higher than the cosmic mean value. 
For the latter mass scale, the boundary is defined as the cloud radius, 
where the ratio of the enclosed mass to the local Jeans mass (Eq. \ref{eq:M_BE}) 
has its maximum value. We present the statistical analyses 
for the cloud properties calculated for the two mass scales.

\begin{figure}
\begin{center}
\resizebox{7.5cm}{!}{\includegraphics[clip,scale=1]{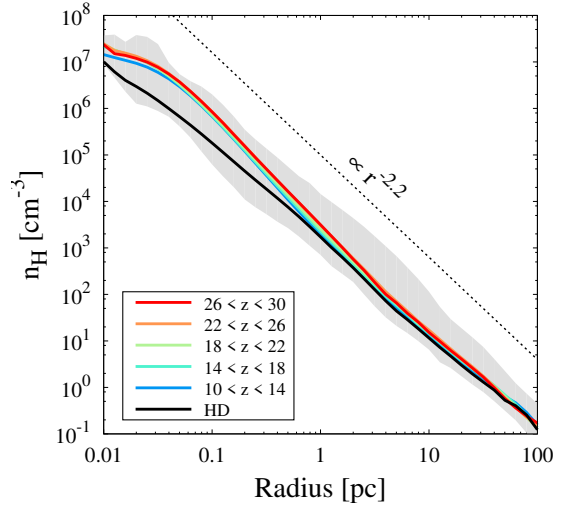}}
\caption{
Radial gas density profiles of primordial star-forming clouds 
when $n_{\rm H,cen} = 10^7 \ {\rm cm^{-3}}$. 
The grey region shows the maximum and minimum for all clouds at each radius. 
The coloured lines show averaged profiles for H$_2$-cooling clouds at five redshift ranges and 
the black line shows the averaged profile for HD-cooling clouds. 
The dotted line shows the power-law slope, $n_{\rm H} \propto r^{-2.2}$ \citep[e.g.,][]{omukai98}.}
\label{fig:plot_Radi-Dens_KPC-N1E07-stack_FULL}
\end{center}
\end{figure}

\begin{figure*}
\begin{center}
\resizebox{15cm}{!}{\begin{tabular}{cc}
\includegraphics[clip,scale=1]{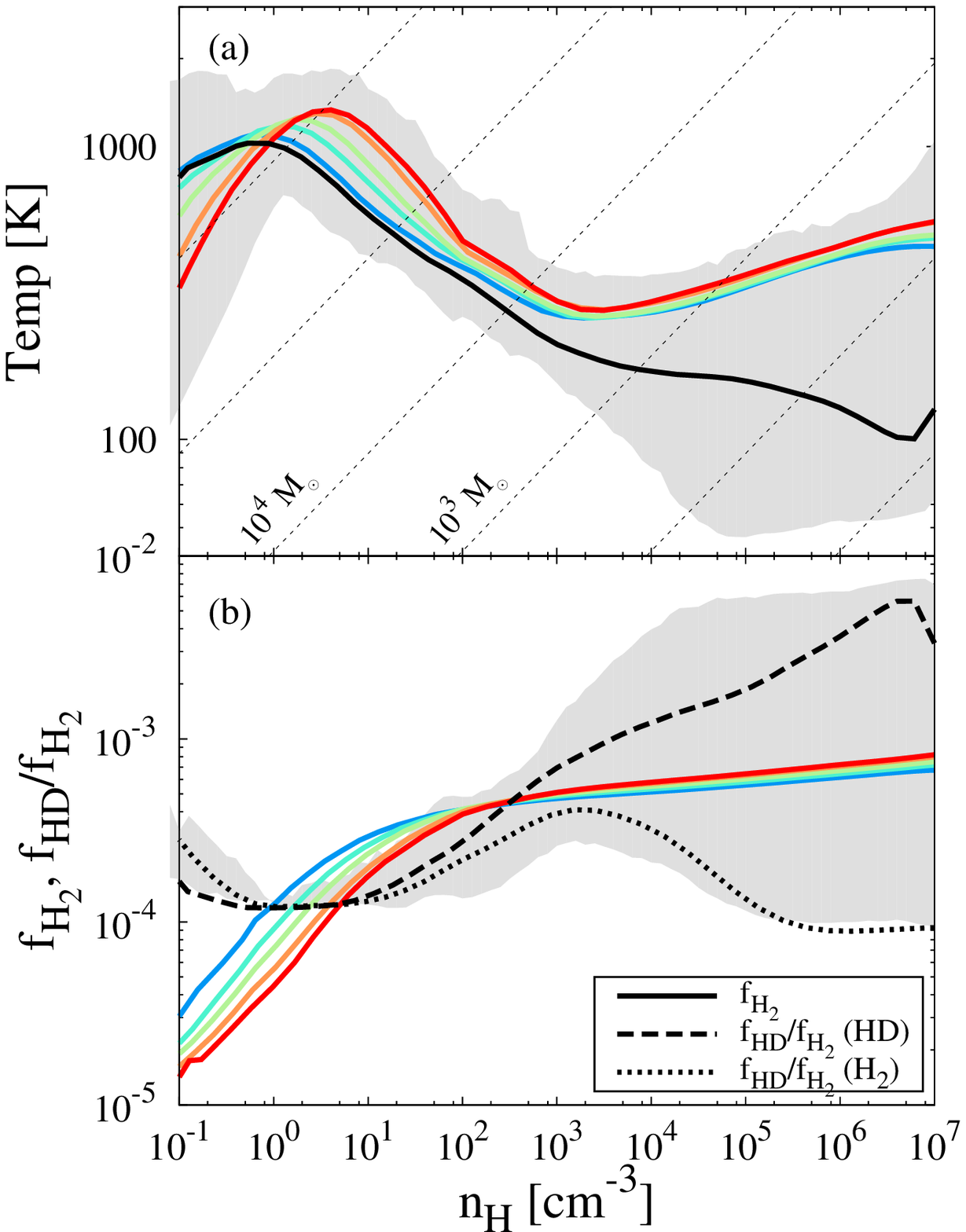}&
\includegraphics[clip,scale=1]{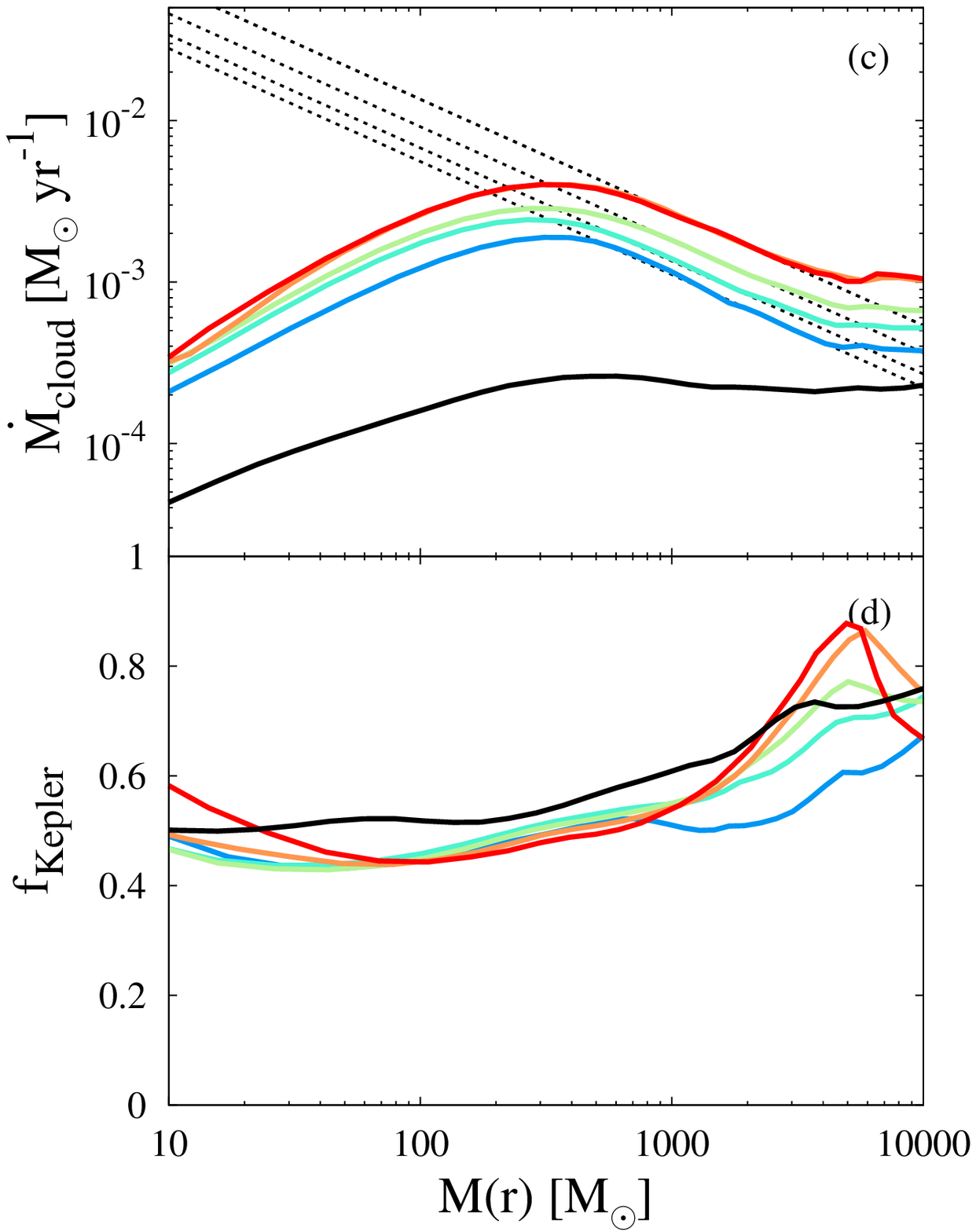}
\end{tabular}}
\caption{
Averaged profiles of primordial star-forming clouds 
when $n_{\rm H,cen} = 10^7 \ \cc$
in the same manner as in Fig. \ref{fig:plot_Radi-Dens_KPC-N1E07-stack_FULL}; 
(a) gas temperature, 
(b) H$_2$ fraction and ratio of HD to H$_2$, 
(c) gas infall rate, and 
(d) Keplerian factor ($f_{\rm Kepler} = v_{\rm rot} / v_{\rm Kepler}$). 
In panel (b), the coloured lines show 
averaged H$_2$ fractions for H$_2$-cooling cases at five redshift ranges. 
The dotted and dashed lines show 
averaged $f_{\rm HD} / f_{\rm H_2}$ for all HD-cooling and H$_2$-cooling cases, respectively. 
In panel (c), the dotted lines show 
the fitting functions for the averaged infall rates at the protostellar formation 
and are characterized by 
$\{$0.34, 0.34, 0.23, 0.17, 0.14$\} \times (M_{\rm enc} / \msun)^{-0.7}$ M$_\odot \ {\rm yr^{-1}}$ 
in order of higher- to lower-redshift groups.
}
\label{fig:plot_KPC-N1E07-stack_FULL}
\end{center}
\end{figure*}

\subsection{Virial Scale: DM Mini-Haloes}


The distributions of the formation redshift and the virial mass of mini-haloes are shown in Fig. \ref{fig:plot_z-Mvir}. 
We see that most of the mini-haloes form at $z = 30 - 10$ with $M_{\rm vir} = 2 \times 10^5 - 1 \times 10^6 \ \msun$. 
The virial temperature of a star-forming halo is about $T_{\rm vir} \simeq 1000$ K~\footnote{This is the critical temperature for a primordial cloud to collapse 
with efficient H$_2$ molecular cooling which is weakly dependent on redshift \citep[e.g.,][]{glover13}. 
In fact, our samples of DM mini-haloes have temperatures close to the critical value (see also Paper I).}, 
and thus the virial mass is
\begin{eqnarray}
M_{\rm virial,3\sigma} \sim 4 \times 10^5 \ \left( \frac{1+z}{20} \right)^{-3/2} \ {\rm M}_{\odot} \; ,
\label{eq:Mvirial_Tvir1000K}
\end{eqnarray}
which gives $M_{\rm vir} = 2.1 \times 10^5$ and $9.9 \times 10^5$ M$_\odot$ at $z = 30$ and 10, 
in good agreement with the typical masses of our mini-haloes. 
As shown in Fig. \ref{fig:plot_z-Mvir}, the average halo mass increases with decreasing redshift, 
which is also consistent with the redshift dependence in Eq. (\ref{eq:Mvirial_Tvir1000K}). 
We have more than 100 haloes per each redshift bin in the range of $z = 22 - 14$, 
which allows us to study even redshift dependences of the properties of our samples.


One of the important quantities at the halo scale is the spin parameter, 
$\lambda \equiv j_{\rm vir}/\sqrt{2} R_{\rm vir} V_{\rm vir}$ 
\citep[following the definition of][]{bullock01}, 
which characterizes the rotational degree of a halo. 
Fig. \ref{fig:plot_2KPC_spin-left+right_half} shows 
the spin parameters for both DM and baryonic components and 
the relative angle of their angular momentum vectors. 
The log-normal distributions and the time (redshift) evolution are consistent with previous studies. 
At high redshift, the baryon spin parameter is lower than that of DM. 
The distribution of baryon spin parameter becomes close to 
that of DM at lower redshift (after $z \sim 14$) 
because of the momentum redistribution between the two components \citep{desouza13}. 
The average angle is $\theta_{\rm ave} \simeq 35^\circ$, suggesting that 
the spin vectors of the two components are roughly aligned with each other in most haloes, 
although with a few exceptions whereby $\theta > 90^\circ$. 
Interestingly, we find a trend that the offset angle and the spin parameter are anticorrelated, 
i.e., the gas and the DM components rotate differently in slowly rotating haloes.

\subsection{Jeans Scale: Gravitationally Unstable Clouds}


In a small halo, the collapsing gas is initially heated adiabatically
until H$_2$ molecular cooling overcomes the compression heating. 
When the H$_2$ fraction reaches the critical value ($f_{\rm H_2,crit} =$ a few $\times 10^{-4}$), 
the temperature begins to decrease rapidly with increasing density.
We follow the collapse until the central hydrogen number density 
$\nhcen$ reaches $10^7~\cc$. 
Among our samples, the time duration of the cloud collapse, 
from  $\nhcen = 10$ to $10^7~\cc$, increases with decreasing 
the formation redshift. 
For instance, the duration averaged over samples for five different
redshift ranges are 29.5~Myr  ($10 < z < 14$), 18.0~Myr ($14 < z < 18$), 
12.0~Myr ($18 < z < 22$), 7.9~Myr ($22 < z < 26$), and 5.5~Myr ($26 < z < 30$).
The clouds collapse more slowly at lower redshifts,
as also found by \citet{gao07}.
In the following, we study the evolution of clouds forming at different redshifts.


In the Pop III.1 case, the thermal evolution proceeds in one of two different modes depending on whether or not HD molecular cooling is effective. 
In most cases, the evolution is driven by H$_2$ cooling alone, whereas HD cooling becomes efficient only under limited conditions. 
In a rapidly rotating cloud, for instance, H$_2$ cooling reduces the temperature to $T \lesssim 200$ K because of the slower collapse and hence less efficient compressional heating. 
HD formation begins at these low temperatures and the resulting HD cooling further reduces the temperature to the cosmic microwave background (CMB) floor at $T_{\rm CMB}(z) \simeq 2.73 \ (1+z)$ K. 
\cite{ripamonti07} study this mode of cloud collapse using 1D hydrodynamic simulations and discuss its overall occurrence rate. 
We also find that this HD-cooling mode occurs in our 3D cosmological simulations (Paper I).\footnote{Another possible route for HD-cooling primordial star-formation is suggested by some studies \citep{uehara00,shchekinov06,prieto12,bovino14b,prieto14}. 
In this scenario, the merging of multiple dark matter haloes induces the formation of shock-waves in which HD formation is enhanced. 
Such merging often yields large mass haloes with $M_{\rm halo} > 10^7 [(1+z)/20]^{-2}$ M$_\odot$ \citep{shchekinov06}, whereas our HD-cooling clouds are found in lower mass haloes with $\sim 5 \times 10^5$ M$_\odot$ at $z \sim 10 - 16$.}


We examine how efficiently HD cooling operates during the collapse in our samples. 
Following previous studies \citep[e.g.,][Paper I]{mcgreer08}, 
we define the HD-cooling case as occurring when the fraction ratio $f_{\rm HD}/f_{\rm H_2}$ 
exceeds $10^{-3}$ at $n_{\rm H,cen} < 10^6 \ {\rm cm^{-3}}$. 
Among our samples, we also find intermediate cases,
whereby the onset of HD cooling is somewhat delayed; i.e.
$f_{\rm HD} / f_{\rm H_2}$ increases sharply in the density range
$10^6 \ \cc < n_{\rm H,cen} < 10^7 \ \cc$ and eventually exceeds $10^{-3}$. 
We classify our 1540 samples of primordial clouds as one of three possible cases,
depending on $f_{\rm HD} / f_{\rm H_2}$ during the collapse: 
H$_2$-cooling cases (1186 cases), HD-cooling cases (151), and the intermediate cases (203), 
where the numbers in parentheses indicate the number of the corresponding samples. 
The H$_2$-cooling cases show a broad range of formation redshifts: 
$10<z<14$ (98), $14<z<18$ (423), $18<z<22$ (474), $22<z<26$ (168), and $26<z<30$ (23), 
which allows us to study the redshift-dependence of the physical properties of the primordial clouds. 
On the other hand, the HD cooling and intermediate cases appear only at low redshift ($z < 18$).


Fig. \ref{fig:plot_Radi-Dens_KPC-N1E07-stack_FULL} shows 
the averaged radial profiles of gas density when $n_{\rm H,cen} = 10^7 \ \cc$. 
They are well fitted by the power-law function 
$\rho \propto r^{-2.2}$ in the envelope \citep[e.g,][]{omukai98}.\footnote{The DM density profiles also represent power-law features 
which at $r \sim 10$ pc transition from $n_{\rm DM} \propto r^{-2}$ to $r^{-1.5}$.} 
The slopes of profiles look very similar, but, for any given radius,
the clouds' densities can differ by more than an order of magnitude (see the grey region). 
Such differences between the clouds will lead to 
differences of the gas infall rates on to a central protostar during the subsequent accretion stage. 
For the H$_2$-cooling clouds, the densities tend to be lower with decreasing formation redshifts. 
The densities of the HD-cooling clouds tend to be even lower than for H$_2$-cooling cases, 
a result of the lower temperature during the collapse as shown below.


Fig. \ref{fig:plot_KPC-N1E07-stack_FULL} displays the averaged thermal (panel a), chemical (panel b), and dynamical (panels c and d) properties of the collapsing clouds. 
Fig. \ref{fig:plot_KPC-N1E07-stack_FULL}(a) shows the temperature distribution as a function of gas density, which reflects the variation of H$_2$ and HD abundances shown in Fig. \ref{fig:plot_KPC-N1E07-stack_FULL}(b). 
We find a small but systematic redshift-dependence of the effectiveness of H$_2$ cooling: lower temperatures for lower redshifts.
This redshift-dependence comes from the fact that the collapse time-scale, which controls the thermal evolution of clouds, is actually longer for the lower redshifts on average. 
The collapse time-scale is set at an early stage of the cloud formation, and thus depends on the cloud's physical properties measured at the virial scale (see Sec.~\ref{ssec:recst_formula} later).
For instance, the mean density at the virial scale is lower for the lower redshifts, which qualitatively explains the slower collapse at the lower redshifts (also see section 4.2.2 in Paper I for a more thorough discussion).
For the HD-cooling cases, the higher $f_{\rm HD}/f_{\rm H_2}$ leads to an even lower temperature. 
The typical mass of a gravitationally unstable cloud is given by the Jeans scale at the local minimum of the temperature profile, which is around ${\rm several} \times 10^3~\msun$ for H$_2$-cooling cases and ${\rm several} \times 10~\msun$ for HD-cooling cases.

Fig. \ref{fig:plot_KPC-N1E07-stack_FULL}(c) shows the radial distributions of the instantaneous gas infall rate $\dot{M}_{\rm cloud} = 4 \pi r^2 \rho v_{\rm rad}$ as a function of enclosed gas mass, extending well outside the Jeans mass.
For the H$_2$-cooling cases, the expected gas infall rates decrease with decreasing formation redshifts. 
For the HD-cooling cases, the infall rates are about 10 times lower. 
We address this issue more in detail in the next subsection. 
The dotted lines show the expected profiles after further collapse. 
We extrapolate the profiles from the outer regions and get power-law profiles $\dot{M}_{\rm cloud} \propto M_{\rm enc}^{0.7}$, which agree with previous results \citep[e.g.,][]{gao07}.

Fig. \ref{fig:plot_KPC-N1E07-stack_FULL}(d) displays the degree of
the azimuthal rotation velocity
$v_{\rm rot}$ normalized by the local Keplerian velocity 
$v_{\rm Kepler} = \sqrt{G M(r) / r}$, where $M(r)$ is the gas mass
within a given radius $r$. 
We calculate $v_{\rm rot}$ by averaging the velocity perpendicular
to the total angular momentum vector for the gas within $r$.
All clouds have similar values clustered around $f_{\rm Kepler} = 0.5$, which agrees with the results of previous studies \citep[e.g.,][]{yoshida06}.

\cite{mckee08} modelled the accretion histories as a
function depending on two parameters: $f_{\rm Kepler}$ and $K^{'}$,
a measure of the entropy of the accreting gas. We can compare their
modelled accretion rates \citep[fig.~9 in][]{mckee08} and our results
 (Fig.~\ref{fig:plot_KPC-N1E07-stack_FULL}c) to determine the
corresponding parameters. By assuming $f_{\rm Kepler} = 0.5$, the other
parameter $K^{'}$ is estimated to be $1 \sim 2$. They also estimate the
final stellar masses when the photoevaporative mass-loss rate overcomes
the accretion rate. The estimated final masses with above parameters
range over $150 - 300~\msun$, which are consistent with our results 
presented in later sections.

\subsection{Distribution of Gas Infall Rates}
\label{ssec:infalldist}


After the onset of collapse, the newly formed protostar grows in mass via accretion from the surrounding envelope. 
It is well known that the accretion history in this phase significantly affects 
the protostar's evolution and the strength of the UV radiative feedback against the accretion flow. 
Since the stellar radiative feedback ultimately shuts off the mass accretion, 
the mean accretion rate is a key parameter that determines the final stellar mass.

\begin{figure}
\begin{center}
\resizebox{8.0cm}{!}{\includegraphics[clip,scale=1]{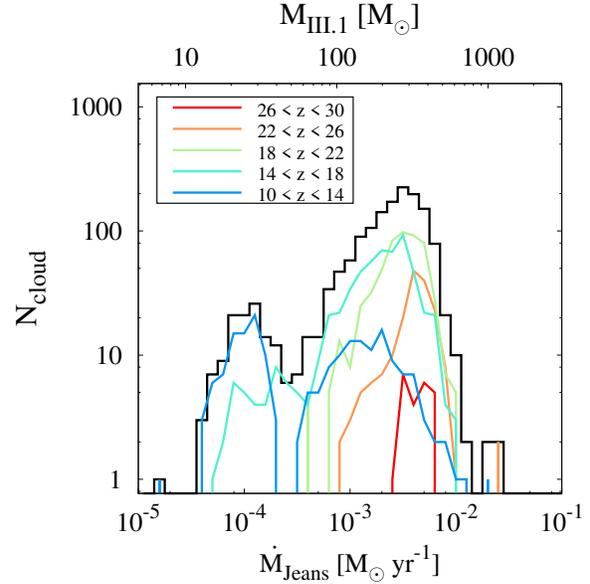}}
\caption{
Histogram of the gas infall rates measured at the Jeans scale 
when $n_{\rm H,cen} = 10^7 \ \cc$. 
The lines with the different colours represent the different redshift ranges 
from high (red) to low redshift (blue).
The top axis gives the stellar mass estimated from 
$\dot{M}_{\rm Jeans}$ by Eq. (\ref{eq:MIII_dMdt-Jeans}).}
\label{fig:plot_dMdt_cloud-Frac}
\end{center}
\end{figure}

Fig. \ref{fig:plot_dMdt_cloud-Frac} shows the histograms of gas infall rates. 
The black line represents the entire collection of samples covering all formation redshifts.
We find two peaks that correspond to the different thermal evolution during the collapse stages: 
the right peak at $\sim 3 \times 10^{-3} \ \msunyr$ is primarily associated with the H$_2$-cooling case and 
the left peak at $\sim 10^{-4} \ \msunyr$ is primarily associated with the HD-cooling case. 
As expected from the redshift-dependence of the thermal evolution, 
the high-rate (high-mass) peak associated with H$_2$-cooling cases is found at all epochs, 
whereas the low-rate peak associated with HD-cooling appears only at low redshifts. 
Thus, the infall rate distribution changes from 
a single peaked function at high redshifts to 
the bimodal function for low redshifts, as shown by the coloured lines.


In Paper I, with $110$ cosmological samples of the primordial star-forming regions, 
we show that the large variation of the gas infall rates leads to a spread of final stellar masses 
ranging from 10 to 1600~$\msun$. 
The results of the present study, with 10 times more samples, 
show the existence of a characteristic value of the infall rate which depends on the formation redshift.
Thus, we can expect that the stellar mass distribution should also have analogous redshift-dependent characteristic values.

\subsection{Estimating Pop III.1 Stellar Masses}
\label{ssec:recst_formula}

\begin{figure}
\begin{center}
\resizebox{7.5cm}{!}{\includegraphics[clip,scale=1]{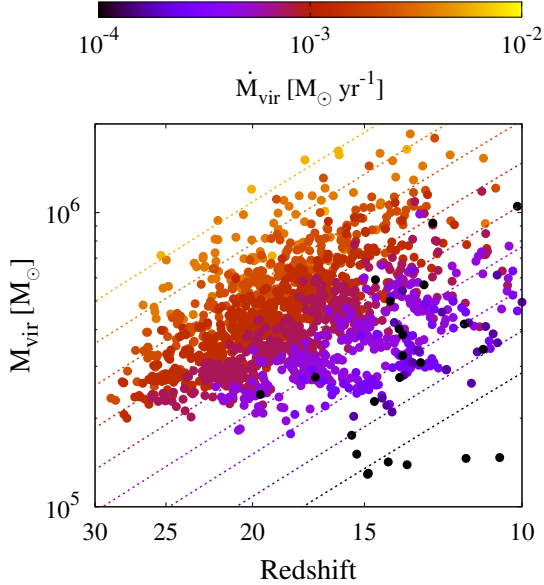}}
\caption{
Variation of the gas infall rate at the virial scale as a function of formation redshift and virial mass. 
The different colours depict the different mass infall rates, according to the colour scale at the top. 
The distributions of the filled circles and dotted lines represent the variation of the sample data and its fitting function Eq. (\ref{eq:dMdt_Virial}). 
}
\label{fig:plor_z-Mvir-dMdtvir}
\end{center}
\end{figure}


Here, we revisit the correlations between the stellar mass $M_{\rm III.1}$
and gas infall rates, which have been examined in Paper I.
We first begin with considering the infall rates measured at the Jeans scale
$\dot{M}_{\rm Jeans}$.
Paper I shows that the gas infall rates measured when $n_{\rm H,cen} = 10^{12} \ {\rm cm^{-3}}$ can be used 
as an accurate estimator, via an empirical fit, of the final stellar masses. 
In the present study, in order to obtain a larger sample at a reasonable computation cost,
we follow the cloud collapse only to $n_{\rm H,cen} = 10^7 \ {\rm cm^{-3}}$. 
We therefore need formulae to predict the Pop III.1 stellar masses 
based on information available at $n_{\rm H,cen} = 10^7 \ {\rm cm^{-3}}$.
To this end, we reanalyse the results of our simulations in Paper I and
determine the correlation between the final stellar masses and the gas infall rates at the Jeans scale 
when $n_{\rm H,cen} = 10^7 \ {\rm cm^{-3}}$. 
In general, the infall rates are lower than those at $n_{\rm H,cen} = 10^{12} \ {\rm cm^{-3}}$, 
but they are still well correlated with the stellar masses (Fig. \ref{fig:plot_III2D_dMdt-Mstar}). 
The new fitting formula is given by
\begin{eqnarray}
M_{\rm III} = 250 \ \msun \left( \frac{\dot{M}_{\rm Jeans}}{2.8 \times 10^{-3} \ {\rm M}_\odot \ {\rm yr^{-1}}} \right)^{0.7} \ .
\label{eq:MIII_dMdt-Jeans}
\end{eqnarray}
The top axis in Fig. \ref{fig:plot_dMdt_cloud-Frac} depicts the stellar mass calculated by this equation. 
In this diagram, the positions of the right and left peaks correspond to $M_* \simeq 264$ and $\simeq 24~\msun$, 
which represent the typical masses of Pop III.1 stars forming via H$_2$-cooling and HD-cooling modes, respectively.


Next, we discuss the dependence of the stellar mass on the halo properties. 
Paper I shows that the gas infall rate at the halo scale $\dot{M}_{\rm vir}$ is also correlated to the stellar mass, 
but the correlation is somewhat weaker. 
Our much larger sample size, however, enables us to derive more reliable correlations than in Paper I. 
We reconstruct the fitting formula at the halo scale, which depends on two parameters: the formation redshift and the halo mass (see Fig. \ref{fig:plor_z-Mvir-dMdtvir}), 
\begin{eqnarray}
\dot{M}_{\rm vir} = 1.1 \times 10^{-3} \, {\rm M}_\odot \, {\rm yr^{-1}} \hspace{-0.8mm}
     \left( \frac{1+z}{20} \right)^{\hspace{-0.7mm}3.5} 
      \hspace{-1mm}
     \left( \frac{M_{\rm vir}}{4 \times 10^5 \ {\rm M}_\odot} \right)^{\hspace{-0.7mm}1.75} 
     \hspace{-2mm} . \hspace{-3mm}
\label{eq:dMdt_Virial}
\end{eqnarray}
We interpret the redshift dependence of the gas infall rates (or the stellar masses) 
shown as the coloured lines in Fig. \ref{fig:plot_dMdt_cloud-Frac} as follows. 
For the H$_2$-cooling cases, the infall rates at the halo scale are well correlated with those at the Jeans scale. 
On the other hand, such a correlation is not found for the HD-cooling cases,
because the infall rates at the Jeans scale are within a narrow range of 
$\dot{M}_{\rm Jeans} \sim 10^{-4}~\msunyr$. 
The relative scaling of $\dot{M}_{\rm vir}$ and $\dot{M}_{\rm Jeans}$ are described by
\begin{eqnarray}
\dot{M}_{\rm Jeans} \hspace{-2.5mm} &= 3 \times \dot{M}_{\rm vir} \qquad   \text{for $\dot{M}_{\rm vir} > 10^{-3}~\msunyr \ ({\rm H_2})$}\, , \\
&= 10^{-4}~\msunyr \ \ \ \qquad      \text{$< 10^{-3}~\msunyr$ (HD)}\, .
\label{eq:dMdt_Cloud-Virial}
\end{eqnarray}
By substituting them into Eq. (\ref{eq:MIII_dMdt-Jeans}), we obtain
\begin{eqnarray}
M_{\rm III.1({H_2})} \hspace{-2mm} &=&\hspace{-2mm}  264 \ \msun \left( \frac{\dot{M}_{\rm vir}}{10^{-3} \ {\rm M}_\odot \ {\rm yr^{-1}}} \right)^{0.7} \ , \label{eq:MIII-dMdt_Virial1} \\
M_{\rm III.1({HD})} \hspace{-2mm} &=&\hspace{-2mm}  24 \ \msun \ . 
\label{eq:MIII-dMdt_Virial}
\end{eqnarray}
Eq. (\ref{eq:MIII-dMdt_Virial1}) for $M_{\rm III.1({H_2})}$ is modified 
by eliminating $\dot{M}_{\rm vir}$ with Eq. (\ref{eq:dMdt_Virial}) as
\begin{eqnarray}
M_{\rm III.1({H_2})} \hspace{-2mm} &=&\hspace{-2mm}  282 \ \msun \left( \frac{1+z}{20} \right)^{2.45} \left( \frac{M_{\rm vir}}{4 \times 10^5 \ {\rm M}_\odot} \right)^{1.23} \hspace{-1mm} .
\label{eq:MIII-Virial}
\end{eqnarray}
Under the assumption that primordial star formation mostly occurs
in $3 \sigma$ mini-haloes with $T_{\rm vir} = 10^3$ K, 
we evaluate the typical value of $M_{\rm III.1({H_2})}$ 
by substituting Eq. (\ref{eq:Mvirial_Tvir1000K}) into (\ref{eq:MIII-Virial}), 
\begin{eqnarray}
M_{\rm III.1({H_2})} \hspace{-2mm} &=&\hspace{-2mm}  282 \ \msun \left( \frac{1+z}{20} \right)^{0.605} \ .
\label{eq:MIII-Virial_3sigma}
\end{eqnarray}
This equation provides the typical stellar mass for the H$_2$-cooling mode 
of $282 \ \msun$ at $1+z = 20$, which is consistent with our results 
shown in Fig. \ref{fig:plot_dMdt_cloud-Frac}. 
Eq. (\ref{eq:MIII-Virial_3sigma}) is also consistent with the fact that, at low redshifts, 
the mass distribution peak moves to lower stellar masses.



\begin{table*}
\begin{center}
\begin{tabular}{lrrrrrrrrrr}
\hline
\hline
ID & \multicolumn{2}{c}{$J_{21}$=0} & \multicolumn{2}{c}{$J_{21}$=0.1} & \multicolumn{2}{c}{$J_{21}$=0.316} & \multicolumn{2}{c}{$J_{21}$=1} & \multicolumn{2}{c}{$J_{21}$=10} \\
& RHD & Estimate & RHD & Estimate & RHD & Estimate & RHD & Estimate &
 RHD & Estimate \\ 
\hline
1 &     | &   $\underline{12.5}$ &     | &  872.6 &     | &  352.3 &  238.1 &  368.3 & \multicolumn{2}{c}{$\times$}  \\ 
2 &   $\underline{15.7}$ &   $\underline{23.7}$ &   89.1 &   93.6 &     | &  286.4 &     | &  449.8 &  183.2 &  203.7  \\ 
3 &   $\underline{25.7}$ &   $\underline{32.1}$ &     | &  150.8 & \multicolumn{2}{c}{$\times$} &     | &  809.5 &     | & 2438.7  \\ 
4 &   $\underline{49.9}$ &   $\underline{49.4}$ &   $\underline{61.0}$ &  $\underline{100.7}$ &     | &  425.7 &  883.3 &  433.7 &  496.1 &  373.6  \\ 
5 &     | &   $\underline{59.1}$ &  152.0 &  317.5 &  177.7 &  236.7 &  126.6 &  176.3 &     | &  503.8  \\ 
6 &  197.2 &  142.2 &   $\underline{40.9}$ &   $\underline{33.0}$ &     -- &  608.2 &   $\underline{29.1}$ &   $\underline{70.6}$ & \multicolumn{2}{c}{$\times$}  \\ 
7 &     | &  328.2 &  235.4 &  268.5 &  175.7 &  544.4 &     | &  522.7 &     | &  473.8  \\ 
8 &     | &  352.3 &   $\underline{60.1}$ &   $\underline{20.8}$ &     | &  397.5 &     | &  655.6 & \multicolumn{2}{c}{$\times$}  \\ 
9 &  291.1 &  384.0 &  323.7 &  406.1 &     | &  367.8 &     | &  593.6 &  618.8 &  176.3  \\ 
\hline
\end{tabular}
\caption{
Column 1: Index of the clouds, 
Columns $2-11$: Final stellar masses in the unit of ${\rm M}_\odot$ for different FUV intensities: $J_{21}$ = 0, 0.1, 0.316, 1, and 10. 
The masses are determined by two different methods: the 2D RHD simulations (left) and from the estimating formula (Eq. \ref{eq:MIII_dMdt-Jeans}; right).
The underbars indicate that the cloud experiences significant HD cooling during the run-away collapse stage. 
The crosses indicate that cloud collapse is prevented by photodissociating radiation.
The dashes indicate the cases without final stellar masses calculated by 2D RHD simulations.
}
\label{tab:M_III2DIS}
\end{center}
\end{table*}

\section{Evaluation of Population III.2$_{\rm D}$ Stellar Masses}
\label{sec:III2D}

\begin{figure}
\begin{center}
\resizebox{7.5cm}{!}{\includegraphics[clip,scale=1]{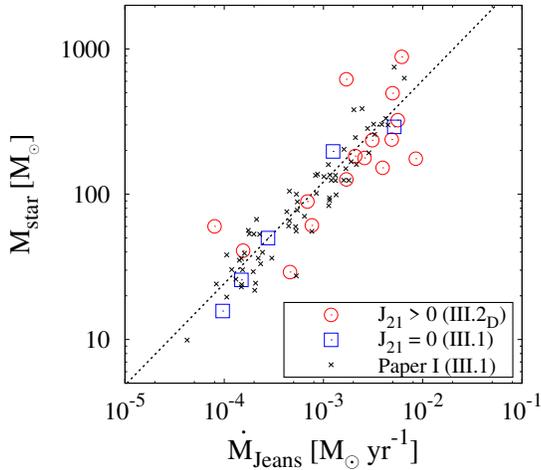}}
\caption{
Final stellar masses obtained in 2D RHD simulations 
as a function of the gas infall rate at the Jeans scale 
when $n_{\rm H,cen} = 10^7 \ {\rm cm^{-3}}$. 
The open circles and squares represent 
16 cases with $J_{21} > 0$ (Pop III.2$_{\rm D}$ stars) and 
5 cases with $J_{21} = 0$ (Pop III.1 stars), 
in which the 2D RHD simulations successfully follow the 
evolution until the final stellar masses are fixed.
The crosses depict the reanalysed results from Paper I
obtained for the same central density. 
The dotted line shows 
our fitting formula for Population III stars (Eq. \ref{eq:MIII_dMdt-Jeans}).}
\label{fig:plot_III2D_dMdt-Mstar}
\end{center}
\end{figure}


In this section, we determine how the mass of a Pop III.2$_{\rm D}$ star 
is affected by external FUV irradiation by performing 45 simulation sets. 
We first conduct 3D hydrodynamical simulations 
to follow the early cloud collapse under different FUV fields up to a central density $n_{\rm H,cen} = 10^{13} \ \cc$. 
The primordial cloud gravitationally collapses in 41 cases out of the total 45 cases. 
In the other 4 cases, gas condensation and the gravitational collapse are prevented 
because H$_2$ molecules are completely photodissociated 
(crosses in Table~\ref{tab:M_III2DIS}).


Then, we switch to 2D RHD simulations to study the mass accretion stage 
until the stellar UV feedback finally shuts off the mass accretion on to the protostar. 
The calculations are finished in 21 cases of the 41 collapsed cases and 
the resultant stellar masses are shown in left-hand columns for each $J_{21}$ 
in Table \ref{tab:M_III2DIS} (HD-cooling clouds are marked by underbars). 
In the remaining 20 cases (dashes in Table~\ref{tab:M_III2DIS}), it was difficult to numerically follow the stellar evolution with time-dependent rapid mass accretion,  
so that their stellar masses are estimated by using the procedure described later in this section.


Fig. \ref{fig:plot_III2D_dMdt-Mstar} shows that 
the final stellar masses obtained from the 2D RHD simulations are well correlated with $\dot{M}_{\rm Jeans}$. 
We find that the correlation for Pop III.2$_{D}$ stars is 
similar to that for Pop III.1 stars in Paper I (Eq. \ref{eq:MIII_dMdt-Jeans}). 
This can be understood by noting that, for the Pop III.2$_{\rm D}$ cases, 
photodissociating radiation affects 
the thermal structure of the envelope
only in the early stage of the collapse.
Once the collapse proceeds and the density increases, 
gas self-shielding becomes effective and 
the subsequent evolution is unaffected by the external radiation. 
The evolution after the birth of the protostar is only modified
with a different accretion history resulting from the different
temperature in the envelope.

Having derived the above correlation, 
we can estimate the stellar mass without computing 
the detailed long-term evolution after the birth of the protostar. 
Instead, we only follow the early collapse stage in 3D 
until the central density reaches $10^7 \ \cc$. 
We determine the final stellar mass for the remaining 20 cases
by using the calculated $\dot{M}_{\rm Jeans}$ and Eq. (\ref{eq:MIII_dMdt-Jeans}) 
which are shown in the right-hand columns for each $J_{21}$ in Table \ref{tab:M_III2DIS}.

\begin{figure*}
\begin{center}
\resizebox{16cm}{!}{\begin{tabular}{c}
\includegraphics[clip,scale=1]{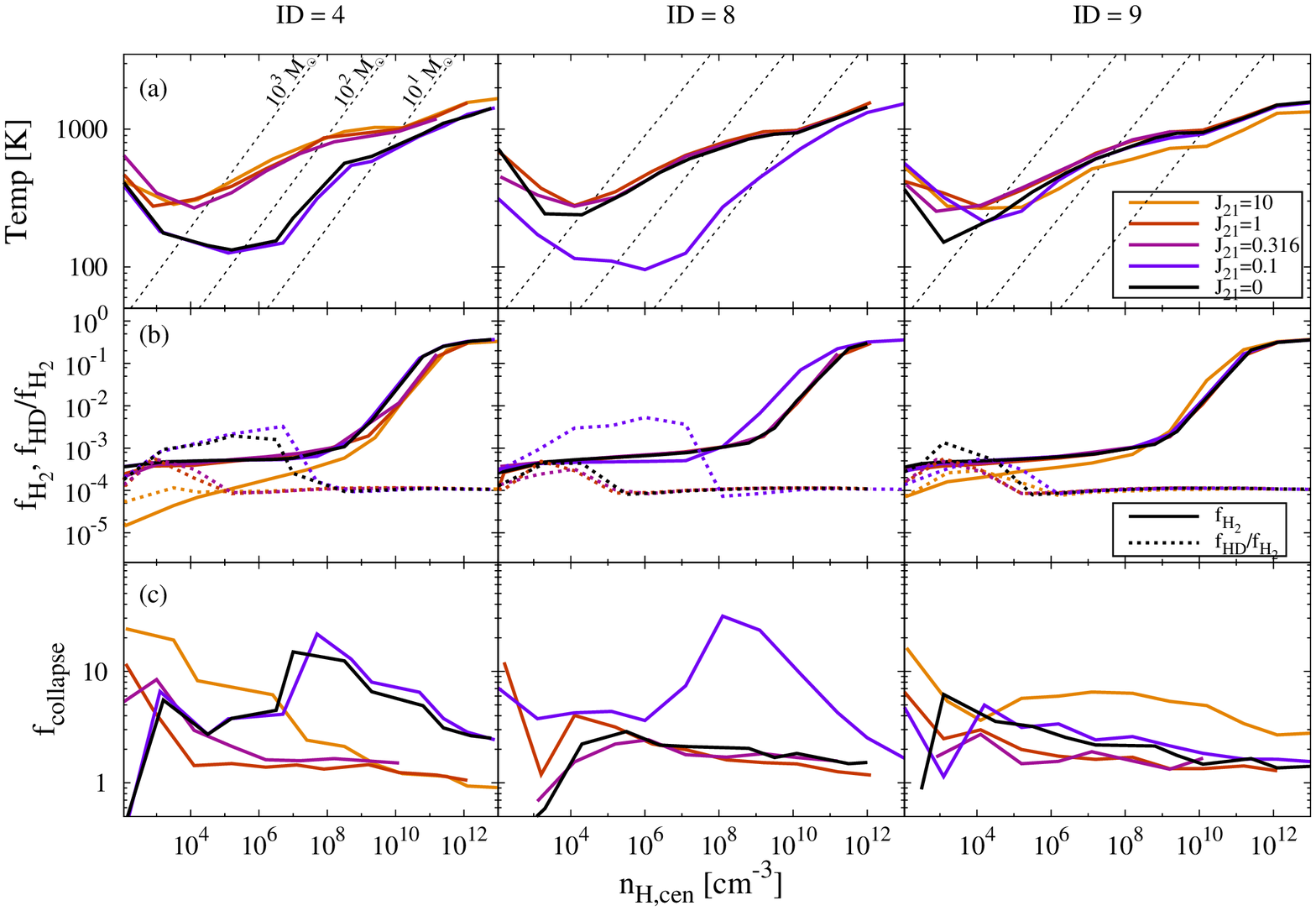}
\vspace{-10mm}\\ \vspace{-5mm}
\includegraphics[clip,scale=1]{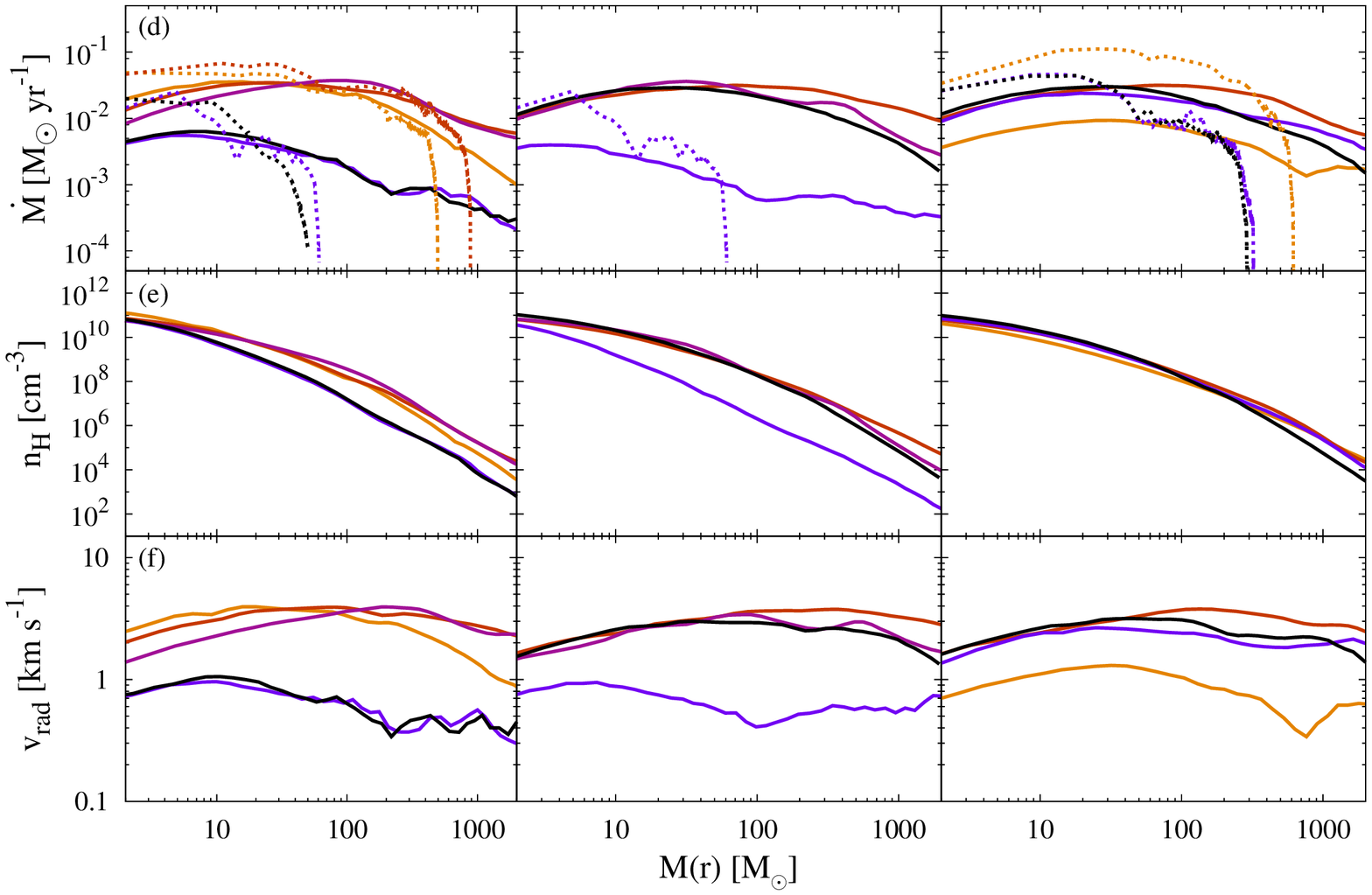}
\end{tabular}}
\caption{
Physical properties of collapsing clouds with different FUV intensity $J_{21}$ 
for the different cases ID = 4 (left), 8 (centre), and 9 (right) listed in Table \ref{tab:M_III2DIS}. 
Line colours in each panel represent 
the different FUV intensity $J_{21}$ = 0, 0.1, 0.316, 1, and 10 from black to yellow. 
The top three rows show the evolution of 
the gas temperature (panels a), chemical abundances (panels b), and 
the ratio between the collapsing time and free-fall time $f_{\rm collapse} = t_{\rm collapse} / t_{\rm ff}$ (panels c) measured at the cloud centre. 
The dotted lines in panel (a) show the Jeans scale as a function of the gas density and temperature. 
In the panels (b), the solid and dotted lines represent 
the number fraction of H$_2$ molecules $f_{\rm H_2}$ and 
the fractional abundance of HD molecules $f_{\rm HD}/f_{\rm H_2}$. 
The bottom three rows show the radial profiles of 
the gas infall rate (panels d), number density (panels e), and radial velocity (panels f) 
as functions of the enclosed mass for $n_{\rm H,cen} = 10^{11} \ {\rm cm^{-3}}$. 
The dotted lines in panels (d) show 
the mass accretion histories seen in the 2D RHD simulations following the evolution after the birth of protostars. 
Note that the horizontal axis represents the protostellar mass for these cases.}
\label{fig:plot_III2-ProfCen_Dens-Temp_multi}
\end{center}
\end{figure*}

\begin{figure*}
\begin{center}
\resizebox{16cm}{!}{\includegraphics[clip,scale=1]{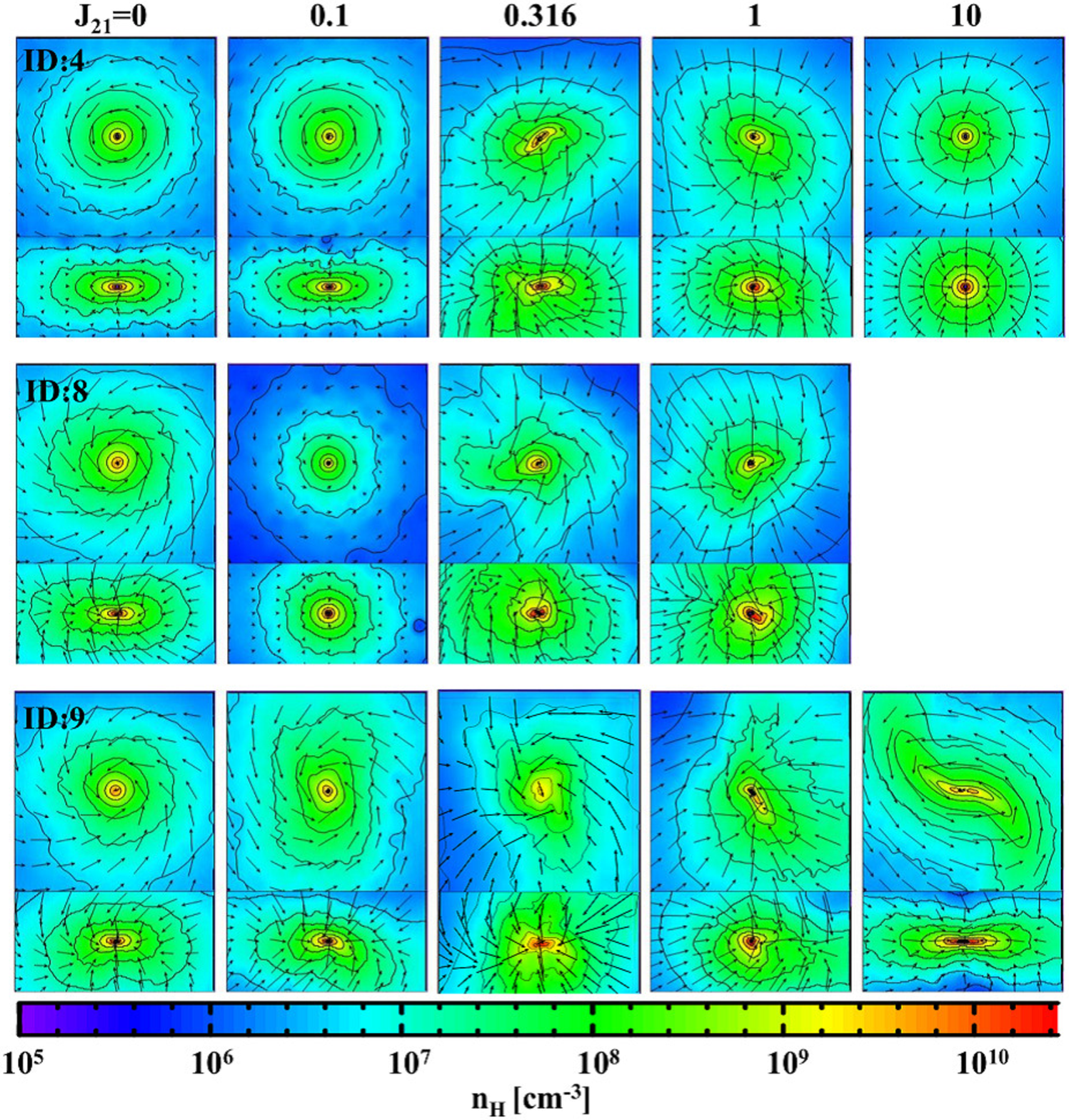}}
\caption{
Density (colour and line contours) and velocity (arrows) distributions around the collapsing centre 
for the same cases as in Fig. \ref{fig:plot_III2-ProfCen_Dens-Temp_multi}. 
The snapshots for $n_{\rm H,cen} = 10^{11} \ {\rm cm^{-3}}$ are presented. 
The box size is 0.1 pc on a side. The top and bottom panels show 
the slices through the $X-Y$ and $X-Z$ planes, whereby the $Z$-axis corresponds to the rotational axis. 
There is a blank panel for ID = 8 with $J_{21} = 10$, 
because the collapse is prevented by the photodissociating radiation in this case.}
\label{fig:III2D_2dmap}
\end{center}
\end{figure*}


We see that, overall, external radiation increases the resulting final stellar masses, 
though the dependence can be quite complex in some cases.
Table \ref{tab:M_III2DIS} shows that the stellar mass does not always increase with increasing FUV intensity. 
For the clouds with ID = 6 and 8, for instance, the stellar mass decreases somewhat from the III.1 value,
when a weak FUV intensity ($J_{21} = 0.1$) is present. 
Interestingly, the thermal evolution during the collapse actually changes 
from an H$_2$-cooling mode to the HD-cooling mode (marked with the overbars) for these cases: i.e.
a weak FUV field sometimes triggers the HD-cooling mode. 
We shall examine the interesting cases in greater detail in the next sections. 
We first review the overall trend of our results, and then examine the detailed evolution for the clouds ID = 4, 8, and 9. 
Finally, we describe how these results are used to derive the mass distributions of Pop III.1 and III.2$_{\rm D}$ stars.

\subsection{General Properties}


Fig. \ref{fig:plot_III2-ProfCen_Dens-Temp_multi} depicts the thermal and dynamical evolution during the cloud collapse. 
The columns represent ID = 4 (left), 8 (middle), or 9 (right). 
For ID = 4 and 8, HD cooling becomes efficient enough to alter the thermal evolution with the weak FUV intensity $J_{21} = 0.1$. 
The decrease of the temperature at $n_{\rm H,cen} \lesssim 10^7 \ \cc$ is explained by 
the increase of HD fraction and hence of the cooling rate (panels a and b). 
As described in Section 4.2, HD cooling becomes efficient when the cloud collapse is slow. 
We calculate the elapsed time, $t_{\rm collapse}$, between two snapshots with $n_{\rm H,cen}$ and $10 \times n_{\rm H,cen}$.
Fig. \ref{fig:plot_III2-ProfCen_Dens-Temp_multi}(c) shows that the time-scale ratio $f_{\rm collapse} \equiv t_{\rm collapse}/t_{\rm ff}$ exceeds 3 in the early stage.\footnote{We have confirmed that the thermal evolution seen in the 3D simulations is well explained with the one-zone models using $f_{\rm collapse}$ as a free parameter (see appendix C in Paper I).}
We also see that the collapse is decelerated around $n_{\rm H,cen} \sim 10^8 \ {\rm cm^{-3}}$, when HD cooling becomes ineffective. 
In addition to the effect of HD cooling, the H$_2$-cooling cases show substantial variations of the temperature, depending on the FUV intensity (for instance, the ID = 9 case).
We conclude that the accretion histories after the birth of a protostar are also dependent on the local FUV radiation field.


The bottom three rows in Fig. \ref{fig:plot_III2-ProfCen_Dens-Temp_multi} show 
the dynamical properties of collapsing clouds at the point in time when $n_{\rm H,cen} = 10^{11} \ \cc$. 
The instantaneous gas infall rate $4 \pi r^2 \rho v_{\rm rad}$ (panel d) depends on the radial distributions of the density (panel e) and velocity (panel f).
The variations seen in the profiles originate from the different thermal evolution during the collapse. 
Fig. \ref{fig:plot_III2-ProfCen_Dens-Temp_multi}(d) also shows that 
the mass accretion rates observed in the 2D RHD simulations (dotted lines), 
which follow the subsequent evolution after the birth of a protostar (the abscissa is now the protostellar mass), 
follow the dependence on FUV intensity expected with the snapshots
before the birth of the protostar (solid lines).


Overall, the photodissociating radiation has a stronger impact on those clouds that evolve on low-temperature tracks 
for $J_{21} = 0$ (see e.g. ID = 4 in Table \ref{tab:M_III2DIS} and Fig. \ref{fig:plot_III2-ProfCen_Dens-Temp_multi}). 
In such cases, the thermal evolutionary path moves towards
higher temperatures with increasing $J_{21}$, 
which results in larger accretion rates and hence ultimately more massive stars. 
The effect is most significant for the clouds that would normally evolve on a HD-cooling path when FUV radiation is absent (e.g., ID = $1 - 5$ cases). 
However, the dependence of the thermal evolution on the FUV intensity is not always simple nor even monotonic in some cases. 
For ID = 8, whose thermal evolution follows the typical H$_2$-cooling mode for $J_{21} = 0$, 
HD cooling dominates when $J_{21}$ is increased to 0.1. 
In the following, we look at individual cases in more detail to understand this complex behaviour.

\subsection{Individual Cases}


To study the evolution of individual cases, 
it is useful to examine the 3D distributions of gas density and velocity for the different FUV intensities. 
Fig. \ref{fig:III2D_2dmap} displays such distributions in the central region with 1 pc on a side. 
We see that, overall, the morphology of a cloud changes from a disc-like 
shape to a more spherical one with increasing $J_{21}$ (from left to right). 
This can be understood as follows. 
At the higher FUV intensity, H$_2$ and HD molecules are photodissociated more efficiently, 
which makes the equation of state (EoS) of the primordial gas stiffer. 
The collapse is decelerated in such a case and there is more time for angular momentum redistribution within the cloud. 
As a result, the cloud becomes more spherical. 
Once the density increases to the point that self-shielding effect becomes important, 
the collapse occurs rapidly with a weaker rotational support. 
This partly explains why the strong FUV irradiation increases the gas infall rate. 
An exceptional case is for ID = 9 with $J_{21} = 10$, 
which has significant rotation, resulting in a disc-like structure with spiral arms. 
We describe the evolution of each case below.

\subsubsection{ID = 4: disabling the HD-cooling path with FUV radiation}


We first consider the case ID = 4, 
for which the cloud evolves on the HD-cooling path for the Pop III.1 case ($J_{21} = 0$). 
Fig. \ref{fig:III2D_2dmap} shows that the cloud rotates rapidly, 
which is also inferred from the low infall velocity shown 
in Fig. \ref{fig:plot_III2-ProfCen_Dens-Temp_multi}(f). 
We remind the reader that the HD-cooling path emerges when the collapse is slow (Paper I), 
which could be the result of rapid rotation. 
The figures show that 
for $J_{21} = 0.1$ the evolution does not change much. 
The photodissociation of molecules hardly affects the evolution because of efficient self-shielding. 
The resulting stellar mass is almost the same as for the Pop III.1 case.


With the higher FUV intensities $J_{21}$ = 0.316 and 1, however, 
the thermal evolution tracks shift from the HD-cooling path to the H$_2$-cooling path,
because of the enhanced photodissociation of H$_2$ and HD molecules. 
The decreased coolants cannot cool the cloud sufficiently to follow a HD-cooling evolution. 
Once the thermal evolution begins to follow the H$_2$-cooling path, 
the cloud becomes more spherical as described above. 
With the reduced rotational support, the collapse proceeds more rapidly and
the gas infall rates are consequently higher than 
for the cases with weaker FUV irradiation $J_{21} \lesssim 0.1$.


The strong FUV irradiation with $J_{21} = 10$ photodissociates H$_2$ molecules 
even up to $n_{\rm H} = 10^9 \ {\rm cm^{-3}}$ (see Fig. \ref{fig:plot_III2-ProfCen_Dens-Temp_multi}b). 
With the reduced abundance of coolants, 
the cloud collapses slowly until three-body H$_2$ formation begins. 
The cloud has a spherical shape as a result of angular momentum redistribution over such a prolonged collapse time. 
The small rotational support results in the highest gas infall rates 
for $M(r) < 100 \ \msun$ (or $n_{\rm H} > 10^9 \ {\rm cm^{-3}}$) among the five cases examined. 
In the outer part of the envelope, however, the infall rate is lower than for $J_{21}$ = 0.316 and 1,
because of the slow collapse before self-shielding becomes effective. 
Due to the low infall rates in the outer low density regions 
(e.g., for $n_{\rm H,cen} = 10^7 \ \cc$, see Fig. \ref{fig:plot_III2-ProfCen_Dens-Temp_multi}d), 
the final stellar mass with $J_{21} = 10$ is lower than with weaker FUV irradiation.


The ID = 4 cases represent the typical
variation of the stellar mass with external photodissociating radiation: 
the stellar mass increases with FUV intensity $J_{21}$.
However, the trend saturates around $J_{21} \sim 10$ and then 
the stellar mass {\it decreases} for even higher FUV intensity. 
Cloud collapse would ultimately be prevented with very strong FUV irradiation, 
which is actually seen in some other cases (crosses in Table \ref{tab:M_III2DIS}).

\begin{figure*}
\begin{center}
\resizebox{14.5cm}{!}{\begin{tabular}{cc}
\includegraphics[clip,scale=1]{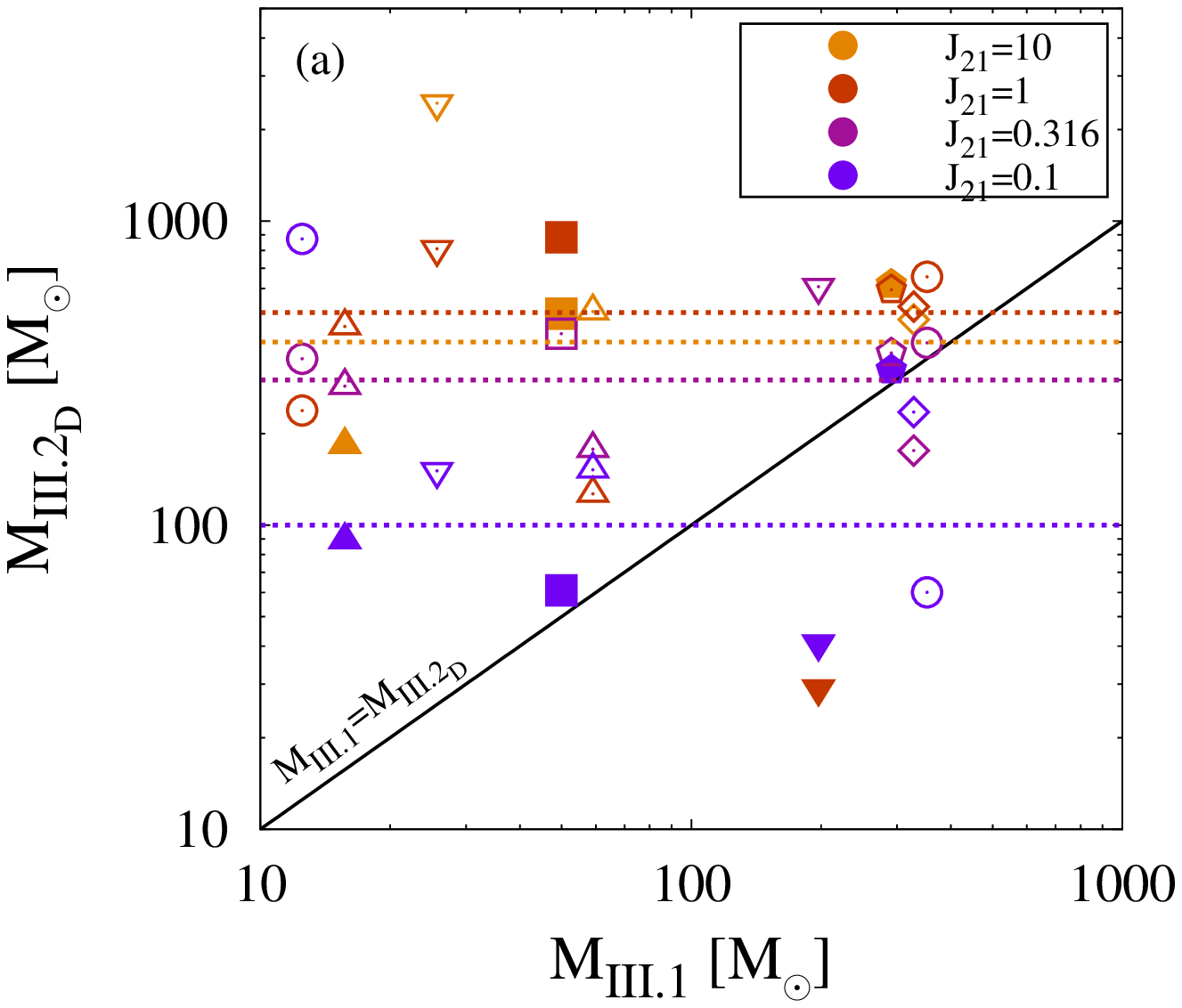}&
\includegraphics[clip,scale=1]{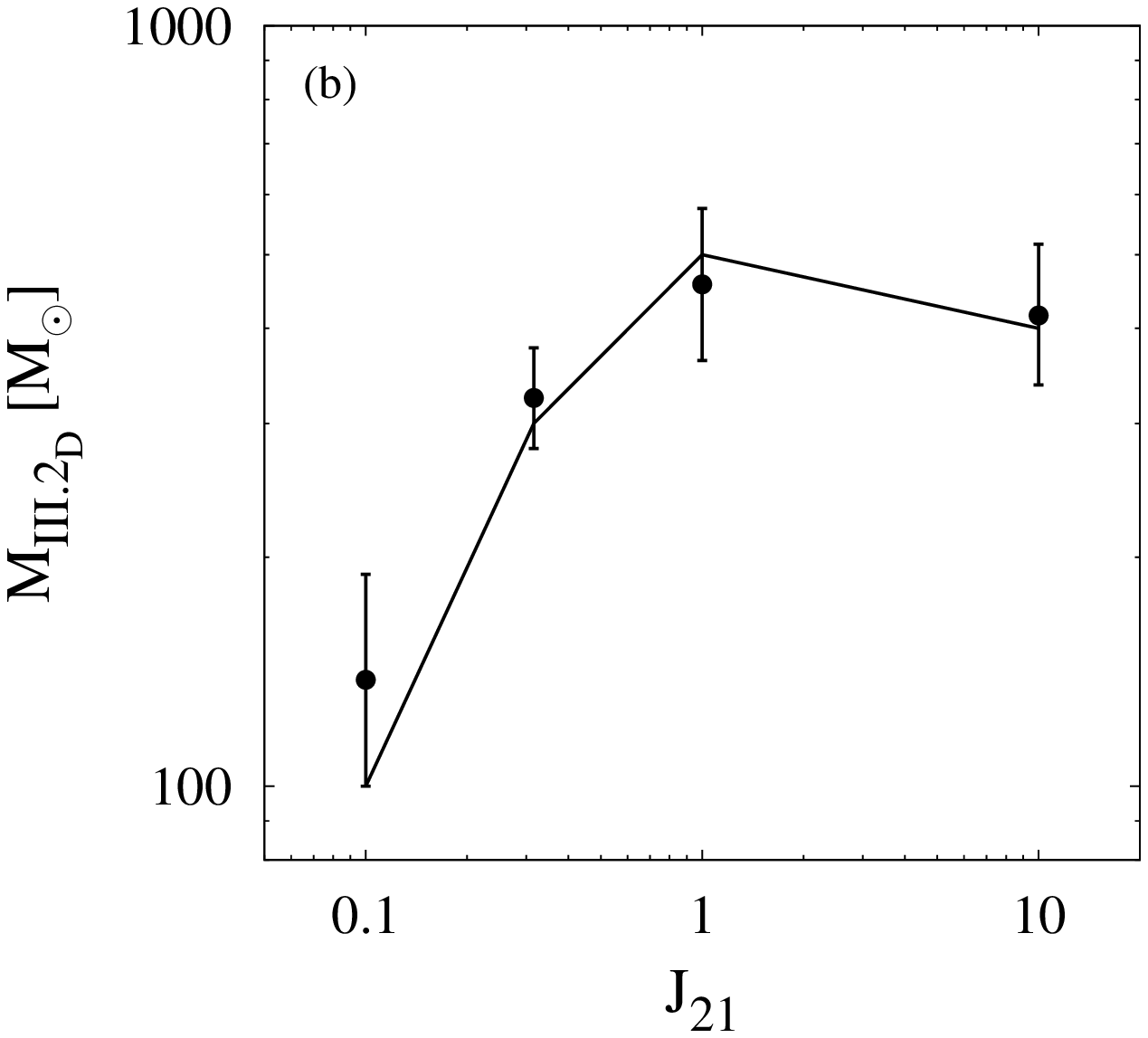}
\end{tabular}}
\caption{
Stellar masses for the Pop III.2$_{\rm D}$ cases with different FUV intensity $J_{21}$. 
The left-hand panel shows the stellar mass of 
Pop III.2$_{\rm D}$ cases $M_{\rm III.2_D}$ ($J_{21} > 0$) plotted versus 
Pop III.1 cases $M_{\rm III.1}$ ($J_{21} = 0$), 
whose different values represent different gas clouds (for ID = $1 \sim 9$ from left to right). 
The filled symbols indicate the stellar masses determined by 2D RHD simulations. 
The values marked with the open symbols, on the other hand, are calculated using
the analytic formula (Eq. \ref{eq:MIII_dMdt-Jeans}). 
The solid line represents the equal mass boundary $M_{\rm III.1} = M_{\rm III.2_D}$. 
The dotted lines show the fitting functions which follow the dependence of $M_{\rm III.2_D}$ on $J_{21}$. 
The right-hand panel shows the correlation of $M_{\rm III.2_D}$ with $J_{21}$.
The symbols with error bars depict the averaged masses and variances of the same $J_{21}$ cases. 
The stellar masses for cases without 2D RHD simulations are given by the estimated values.
The line shows the simplified relation $M_{\rm III.2_D}$ as a function of $J_{21}$
(Eq.~\ref{eq:MIII2D_from_J21}; $M_{\rm III.2_D} = 100$, 300, 500, and 400~M$_\odot$ for $J_{21} = 0.1$, 0.316, 1, and 10),
which is used for evaluating Pop III.2$_{\rm D}$ stellar masses.}
\label{fig:plot_III2D_M31-Rate}
\end{center}
\end{figure*}

\subsubsection{ID = 8: triggering the HD-cooling path with weak FUV radiation}


The next cases we examine in detail are for ID = 8, whereby for $J_{21} =0$ the cloud collapses along the H$_2$-cooling path. 
Interestingly, the HD-cooling path appears only for the weak FUV irradiation $J_{21} = 0.1$ in this case. 
The main effect of the FUV radiation is photodissociating molecules. 
Without molecules, the EoS becomes stiffer and the gas collapses slowly.  
However, the slow collapse actually promotes the formation of H$_2$ and HD molecules. 
Enhanced self-shielding eventually stops further photodissociation, 
triggering the HD-cooling mode in the subsequent evolution. 
At higher intensities $J_{21}$ = 0.316 and 1, 
the thermal evolution tracks return to the H$_2$-cooling path,
because the self-shielding is insufficient to prevent photodissociation. 
With $J_{21}$ = 10, the cloud does not collapse even after $10^8$ yr. 
The strong photodissociation completely quenches star formation in this case. 
Our results suggest that 
the critical FUV intensity for preventing collapse depends on 
the physical properties of the cloud such as its morphology and the degree of rotation. 
This is likely, because the strength of the self-shielding effect depends on 
the radial profiles of the density within the cloud.

\subsubsection{ID = 9: reducing temperature with strong FUV
   radiation for the H$_2$-cooling path}


Finally, we consider ID = 9 cases, which show an exceptional behaviour
for strong FUV irradiation.
The gas infall rate and resulting final stellar mass
monotonically increase with increasing $J_{21}$ for $J_{21} \leq 1$.
For $J_{21}$ = 10, however, in spite of the intense FUV irradiation, 
the cloud evolves through the lowest temperature path among all cases examined
(Fig. \ref{fig:plot_III2-ProfCen_Dens-Temp_multi}a). 
This is because the strong photodissociation decelerates the collapse 
even after the density becomes very high. 
In fact, Fig. \ref{fig:plot_III2-ProfCen_Dens-Temp_multi}(c) shows that 
the time-scale ratio $f_{\rm collapse}$ is around 5 at $n_{\rm H} \gtrsim 10^9 \ \cc$. 
Paper I finds that once three-body H$_2$ formation becomes effective, 
the gas temperature is lower for the longer collapse time-scales (see their appendix C);
i.e., more H$_2$ molecules, the coolant, are formed with the slower collapse. 
The enhanced coolant abundance results in a softer EoS, 
which in turn leads to a more flattened morphology as the collapse advances \citep{hanawa00}.
In fact, the cloud morphology accordingly changes from disc-like 
to a more spherical structure for the cases with $J_{21} \leq 1$,
but shows the highly rotating disc-like shape for
$J_{21} = 10$ as seen in Fig. \ref{fig:III2D_2dmap}.

\subsection{Formula for Estimating Population III.2$_{\rm D}$ Stellar Masses}


We have seen that the masses of Pop III.2$_{\rm D}$ stars $M_{\rm III.2_D}$ vary not only among different gas clouds but also by the effect of FUV irradiation.
Despite the fact that $M_{\rm III.2_D}$ is dependent on multiple parameters in a complicated way, the overall trend can be modelled by using the numerical simulation results obtained in this paper.
Fig. \ref{fig:plot_III2D_M31-Rate}(a) summarizes how the stellar mass $M_{\rm III.2_D}$ varies with the parameters ($M_{\rm III.1}$ and $J_{21}$). 
Note that, in the figure, the different Pop III.1 stellar masses indicate that the gas clouds are different.
Overall, the mass growth rate $M_{\rm III.2_D} / M_{\rm III.1}$ increases with decreasing $M_{\rm III.1}$ for a given FUV intensity; gas clouds with low temperatures are more susceptible to FUV irradiation.
On the other hand, the mass increase is relatively small for the cases with the higher Pop III.1 masses, i.e., $M_{\rm III.1} \gtrsim 100 \ \msun$.
In Fig. \ref{fig:plot_III2D_M31-Rate}(a), we see that the same colours for a given $J_{21}$ are distributed around the similar values of $M_{\rm III.2_D}$. 
The Pop III.2$_{\rm D}$ stellar mass does not change radically within a given FUV intensity $J_{21}$. 
Table~\ref{tab:M_III2DIS} shows the diversity of stellar masses for a given $J_{21}$ over a factor of 10 but there is also a systematic dependence.
Fig. \ref{fig:plot_III2D_M31-Rate}(b) shows the averaged values and variances for given values of $J_{21}$.
To obtain the averaged values and its deviations, we use
the estimated stellar masses for the cases without the 2D RHD results.
There is a general trend that $M_{\rm III.2_D}$ increases for $0.1 \leq J_{21} \leq 1$ but {\it decreases} gradually for $J_{21} \geq 1$.
We model such variations of $M_{\rm III.2_D}$ for different $J_{21}$ by the solid line as
\begin{eqnarray}
M_{\rm III.2_D}(J_{21}) = 
  \begin{cases}
     900 \cdot 10^{+0.96 x} \hspace{1.0mm} \left( {\rm if}~\hspace{2.8mm} -1 < x < -0.5 \right), \\
     500 \cdot 10^{+0.44 x} \hspace{1.0mm} \left( {\rm if}~-0.5 < x < 0 \right), \\
     500 \cdot 10^{-0.10 x} \hspace{1.0mm} \left( {\rm if}~\hspace{6.2mm} 0 < x < 1 \right), \\
     400 \hspace{14.9mm} \left( {\rm if}~\hspace{6.0mm} 1 < x \right), 
  \end{cases}\hspace{-10mm}
\label{eq:MIII2D_from_J21}
\end{eqnarray}
where $x = \log_{10} (J_{21})$.

\section{Mass Distribution of Primordial Stars}
\label{sec:distribution}


In this section, we calculate the mass distribution of Pop III.1 and III.2$_{\rm D}$ stars using the non-biased cosmological samples of the primordial star-forming regions (Sec. \ref{sec:sampling}). 
We determine the stellar mass for each halo by considering the physical properties of the cloud for Pop III.1 cases (Eq.~\ref{eq:MIII_dMdt-Jeans}) and intensity of the local FUV radiation field for Pop III.2$_{\rm D}$ cases (Eq.~\ref{eq:MIII2D_from_J21}).

\subsection{Classification into Population III.1 and III.2$_{\bf D}$ Stars}
\label{sec:distribution_ssec:classification}


In the cosmological context, as shown in Fig. \ref{fig:plot_pre-zoom}, 
the earliest generation of stars are formed in filaments or knots of the large scale structure and 
hence the star-forming regions are distributed in a biased manner. 
Radiation emitted by stars formed early affects subsequent star formation in nearby regions. 
As described in Section \ref{ssec:j21calc}, 
we evaluate the local FUV intensity at each cloud considering the spatial distribution of other stars at 
the moment when the cloud central density reaches $10~\cc$. 
We assume that Pop III.2$_{\rm D}$ stars form, when $J_{21} > 0.1$. 
Fig. \ref{fig:J21MAP} shows the resultant FUV radiation fields in the cosmic volume at five different redshifts. 
The mean FUV intensity decreases with decreasing redshift 
mostly because spatial separations between stars are stretched by cosmic expansion. 
At $z = 25$, there is a small cluster of Pop III.2$_{\rm D}$ stars 
that are formed under the influence of a strong FUV field. 
Once the bright massive stars form in a dense region, 
the resulting FUV field could trigger the sequential formation of Pop III.2$_{\rm D}$ stars in nearby haloes. 
At $z = 15$, on the other hand, there are no Pop III.2$_{\rm D}$ stars 
in the simulation volume because of the weak FUV radiation field resulting from
the large physical separation between star-forming regions.


\begin{figure}
\begin{center}
\resizebox{7.5cm}{!}{\includegraphics[clip,scale=1]{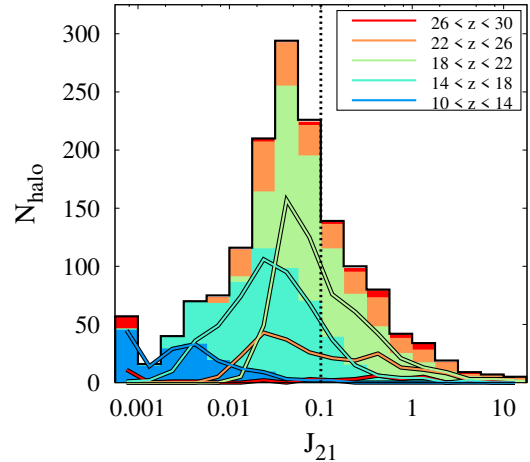}}
\caption{
Histogram of the normalized intensity of photodissociating radiation $J_{21}$ 
at the position of each primordial cloud when $n_{\rm H,cen} = 10 \ {\rm cm^{-3}}$. 
The different colours depict the same redshift ranges as in the right-hand panel
of Fig. \ref{fig:plot_z-Mvir}.
The dotted vertical line represents the critical value of
$J_{21} = 0.1$, above which Pop III.2$_{\rm D}$ stars form.
The haloes with $J_{21} > 10$ and $J_{21} < 10^{-3}$ are included in
the rightmost and leftmost bins, respectively.
}
\label{fig:plot_z-nJ21}
\end{center}
\end{figure}

\begin{figure*}
\begin{center}
\resizebox{15cm}{!}{\begin{tabular}{cc}
\includegraphics[clip,scale=1]{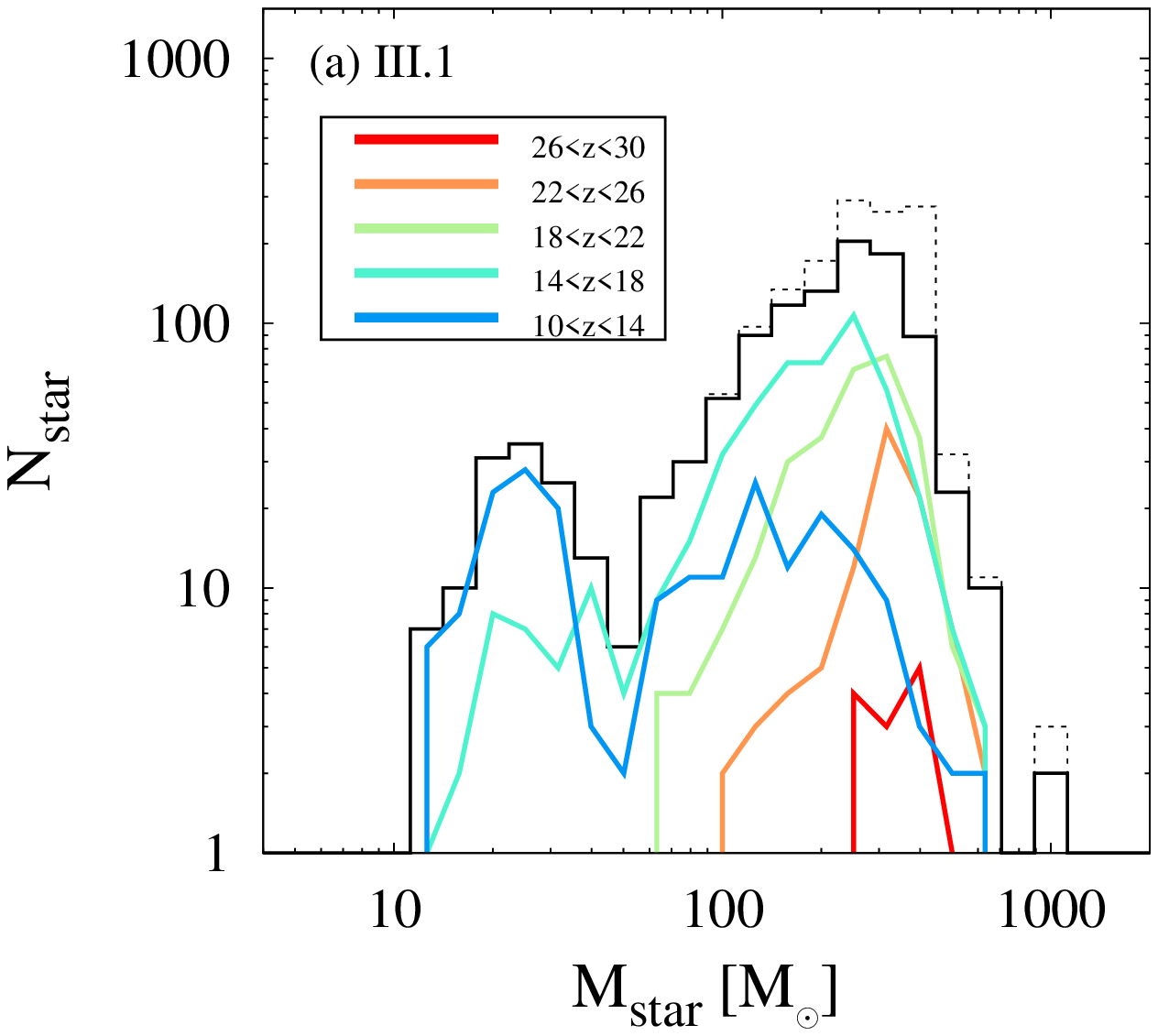}&
\hspace{-15mm}
\includegraphics[clip,scale=1]{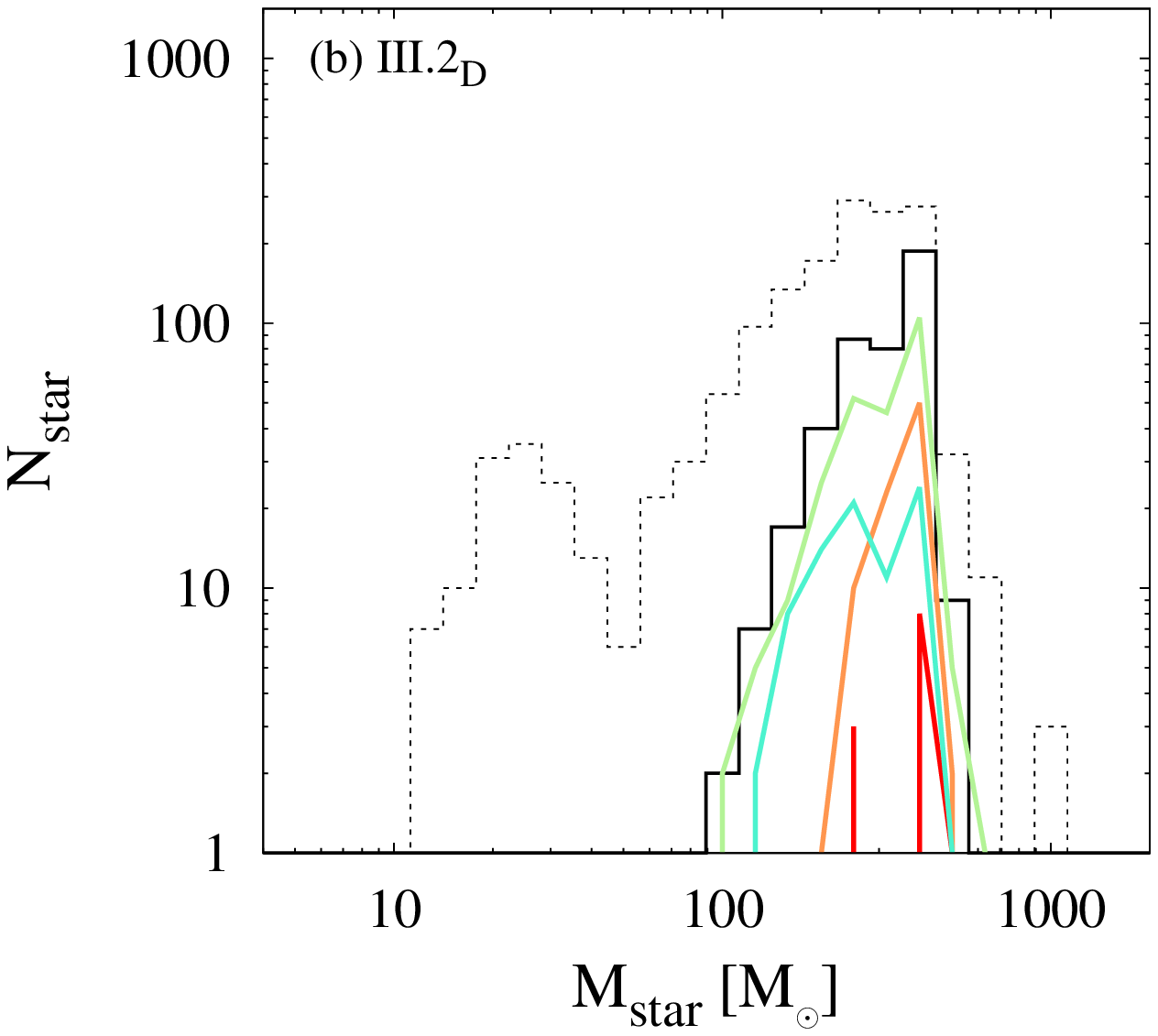}
\end{tabular}}
\caption{Resultant mass distributions of 
Pop III.1 (left) and III.2$_{\rm D}$ (right) stars for the different redshifts. 
The different colours represent the same redshift ranges 
as in the right-hand panel of Fig. \ref{fig:plot_z-Mvir}.
The black solid lines show the total distributions over all redshifts for each population whereas the dotted lines show the sum of them.
}
\label{fig:plot_IMF_dz}
\end{center}
\end{figure*}

Fig. \ref{fig:plot_z-nJ21} shows 
the histogram of the calculated FUV intensity and its redshift dependence. 
About one third of haloes is irradiated by a strong FUV field and thus satisfies our criterion for Pop III.2$_{\rm D}$ star formation. 
At $z \gtrsim 20$, the clouds have a large variation of $J_{21}$, ranging from 0.01 to 10;
about half of the clouds are exposed to FUV radiation with $J_{21} > 0.1$. 
At $z < 20$, however, the fraction of III.2$_{\rm D}$ stars rapidly decreases. 
The transition is caused by a combination of
the increase of physical separations between the clouds, the decrease of the number of FUV-active stars, 
and the systematic decrease of stellar masses by the promoted HD-cooling mode of star-formation. 
We also find that, at $15 < z < 25$, almost all the clouds have non-zero intensity $J_{21} > 0.01$. 
However, such a weak FUV irradiation causes negligible effects on the cloud collapse.

Regarding clouds under strong FUV irradiation with $J_{21} > 10$, we
have 5 such samples at $z = 21 - 26$ with intensities of $J_{21} =$ 12.1,
14.1, 17.7, 70.3, and 97.3. However, these are still below
the critical value of $J_{21}^{\rm crit} \propto \mathcal{O}(1000)$,
above which the so-called direct collapse might be triggered in
atomic-cooling haloes \citep[e.g.,][]{wolcott-green11a}. 
In the current cases with the weaker FUV radiation and lower mass
haloes, the gas clouds likely yield massive Pop III.2$_{\rm D}$ stars
but not the massive BHs as prospected in the direct collapse scenario.

In the next section, we determine the stellar mass for each of 1540 star-forming clouds according to the star-formation mode: 
we use Eq. (\ref{eq:MIII_dMdt-Jeans}) for Pop III.1 stars and 
Eq. (\ref{eq:MIII2D_from_J21}) for Pop III.2$_{\rm D}$ stars.

\subsection{Mass Distribution of Primordial Stars}


Fig. \ref{fig:plot_IMF_dz} displays the mass distributions of Pop III.1 and III.2$_{\rm D}$ stars. 
The black solid lines show the mass distributions integrated over redshifts. 
As expected from Fig. \ref{fig:plot_dMdt_cloud-Frac}, 
the Pop III.1 mass distribution has two peaks around $M_* \simeq 250$ and $25 \ \msun$, 
which reflect the contributions of the H$_2$-cooling and the HD-cooling modes (see Sec. \ref{ssec:infalldist}). 
The mass distribution of Pop III.2$_{\rm D}$ stars is shifted to larger masses $M_* \simeq 400 \ \msun$. 
We find a wide mass range for each population:
$50 \lesssim M_{\rm III.1_{H_2}}/{\rm M}_\odot \lesssim 1000$, 
$10 \lesssim M_{\rm III.1_{HD}}/{\rm M}_\odot \lesssim 50$, and 
$100 \lesssim M_{\rm III.2_{D}}/{\rm M}_\odot \lesssim 1000$, respectively. 
About a half of $M_* > 200 \ \msun$ stars are Pop III.2$_{\rm D}$ stars.
The mass distributions for the different redshifts are represented by lines with different colours. 
For Pop III.1 cases, as suggested by Eq. (\ref{eq:MIII-Virial_3sigma}), 
the high-mass peak, corresponding to the H$_2$-cooling mode, gradually shifts to lower stellar masses with decreasing redshift; 
from $375 \ \msun$ at $z = 30$ to $191 \ \msun$ at $z = 10$. 
The mass distribution of Pop III.2$_{\rm D}$ stars shows a similar trend, 
which is caused by the decrease of the average FUV intensity as shown in Fig. \ref{fig:plot_z-nJ21}. 
We find no such shift for the HD-cooling cases, probably because of the small sample size at lower redshifts. 
Overall, the decrease of the Pop III.2$_{\rm D}$ fraction, 
the decrease of Pop III.1$_{\rm H_2}$ stellar masses, and 
the increase of low-mass Pop III.1$_{\rm HD}$ fraction 
explain the shift of the distribution to the lower masses with decreasing redshift.


Because we use the analytic function Eq. (\ref{eq:MIII_dMdt-Jeans}) 
to estimate the Pop III.1 stellar masses from the gas infall rates at the Jeans scale, 
the mass distribution reflects that of the infall rates. 
Interestingly, except for Pop III.1 cases formed via the HD-cooling mode, 
the stellar mass distributions are well described by power-law functions at both the low-mass and high-mass ends.
For example, the mass distribution of Pop III.1$_{\rm H_2}$ stars is approximately proportional to
$M_*^{2.5}$ at the low-mass end and to
$M_*^{-4}$ at the high-mass end. 
This allows us to define the normalized stellar mass function, 
\begin{eqnarray}
\Psi(M_*) &=& \frac{1.62}{M_{\rm p} / {\rm M}_\odot} 
\left( \frac{M_*}{M_{\rm p}} \right)^{2.5} \ \text{for $M_* < M_{\rm p}$} \ , \\ 
&=& 
\frac{1.62}{M_{\rm p} / {\rm M}_\odot} 
\left( \frac{M_*}{M_{\rm p}} \right)^{-4} \ \text{for $M_* > M_{\rm p}$} \ ,
\label{eq:IMF_III1_H2}
\end{eqnarray}
where $M_{\rm p}$ is the peak mass given by Eq. (\ref{eq:MIII-Virial_3sigma}) and 
$\Psi (M_*)$ is normalized by $\int \Psi (M_*) \ dM_* = 1$.
This is obviously quite different from the well-known Salpeter function. 
The above equation allows us to model the time-dependent mass function of primordial stars.

\begin{figure}
\begin{center}
\resizebox{7.5cm}{!}{\includegraphics[clip,scale=1]{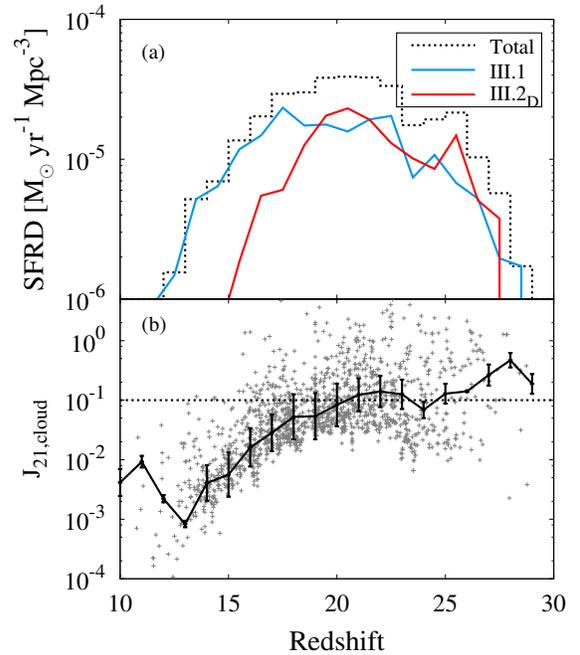}}
\caption{
Redshift evolution of SFRD of Population III.1 and III.2$_{\rm D}$ stars (panel a) and the averaged $J_{21}$ for clouds with its variance (panel b). 
In panel (a), the dotted line represents the total of them.
In panel (b), the grey dots show the scatter of $J_{21}$ at each cloud and the dotted line represents the critical value of $J_{21} = 0.1$, above which Pop III.2$_{\rm D}$ stars form.
}
\label{fig:plot_z-SFRD}
\end{center}
\end{figure}

\begin{table*}
\begin{center}
\begin{tabular}{ccrrrrrrrrrrr}
\hline
\hline
Fate & Mass Range & III.1 & III.$2_{\rm D}$ & & & & & & & & & Total \\ 
 & (M$_\odot$) & \multicolumn{2}{c}{($10<z<14$)} & \multicolumn{2}{c}{($14<z<18$)} & \multicolumn{2}{c}{($18<z<22$)} & \multicolumn{2}{c}{($22<z<26$)} & \multicolumn{2}{c}{($26<z<30$)}  & \\
\hline
NS & \ \ \ \ \ \ \ \ \ $M_*< \ 25$ &  2.4\% &  &  0.7\% &  &  &  &  &  &  & & 3.1\% \\
LMBH & \ \ $25<M_*< \ 80$ &  4.2\% &  &  2.3\% &  &  0.3\% &  &  &  &  & & 6.8\% \\
PPISN & \ \ $80<M_*<120$ &  1.4\% &  &  2.7\% &  &  0.7\% &  0.1\% &  0.1\% &  &  & & 5.0\% \\
PISN & $120<M_*<240$ &  3.7\% &  & 12.1\% &  1.4\% &  4.8\% &  2.2\% &  0.7\% &  0.1\% &  & & 25.0\% \\
HMBH & $240<M_*$ \ \ \ \ \ \ \ \ &  2.0\% &  0.1\% & 13.8\% &  4.0\% & 13.0\% & 14.2\% &  5.6\% &  5.7\% &  0.9\% & 0.8\% & 60.1\% \\
\\
Total & & 13.7\% & 0.1\% & 31.6\% & 5.4\% & 18.8\% & 16.5\% & 6.4\% & 5.8\% & 0.9\% & 0.8\% & 100.0\% \\
\hline
\end{tabular}
\caption{Column 1: Final fate of stellar evolution: neutron star (NS), low-mass black hole (LMBH), pulsation pair-instability supernovae (PPISN), pair-instability supernovae (PISN), and high-mass black hole (HMBH), Column 2: Corresponding mass ranges at ZAMS for each fate from fig. 12 in \citet{yoon12} for the non-rotating case, Columns 3-12: Fractions of each fate in the sample for each redshift range, Column 13: Sum over each row.}
\label{tab:FinalFate}
\end{center}
\end{table*}

\subsection{Star Formation Rate Density}


Fig. \ref{fig:plot_z-SFRD}(a) shows the star formation rate densities (SFRD) as a function of redshift. 
The primordial SFRD rises until $z \sim 20$ and decreases afterwards 
(as Pop II star formation becomes the main mode). 
This evolution of the total SFRD is consistent with previous studies \citep[e.g.,][]{agarwal12,johnson13}, 
but our results show clearly the contribution of Pop III.2$_{\rm D}$ stars. 
Significant Pop III.2$_{\rm D}$ star formation occurs after Pop III.1 stars are formed and emit copious amounts of FUV photons. 
Remarkably, ${\rm SFRD_{III.2_{D}}}$ approaches the same level as ${\rm SFRD_{III.1}}$ at $z \gtrsim 20$.
At $z \lesssim 20$, the fraction of Pop III.2$_{\rm D}$ stars rapidly decreases as shown in Fig. \ref{fig:plot_z-nJ21}. 
Note, however, we may be underestimating ${\rm SFRD_{III.2_{D}}}$ at this point, 
because we ignore the contributions of Pop II stars to the local and global FUV radiation field. 
The FUV background radiation intensity may exceed $J_{21} = 0.1$ at $z \lesssim 10$ and 
the Pop II SFRD may dominate at $z \lesssim 15$ \citep[e.g.,][]{agarwal12}. 
If a large number of Pop II stars are formed during the epoch we consider here, 
the number fraction of Pop III.2$_{\rm D}$ stars could be enhanced and the mass distribution at $z \lesssim 15$ may be significantly modified. 
There are a number of $M_* < 100 \ \msun$ Pop III.1 stars forming via the HD-cooling mode, 
which is easily changed to the H$_2$-cooling mode even for even weak FUV fields (see Sec. \ref{sec:III2D}). 
The enhancement of the stellar mass by photodissociating molecules is significant for these cases (Fig. \ref{fig:plot_III2D_M31-Rate}). 
However, because the total primordial SFRD decreases for $z \lesssim 20$, 
the uncertainties described above would not greatly change the overall mass distribution integrated over redshifts.

Fig. \ref{fig:plot_z-SFRD}(b) shows the redshift evolution of averaged local FUV intensity at each cloud. 
The averaged value decreases with decreasing redshift and falls below
the critical value at $z \sim 20$, which is consistent with the decline
of SFRD for Pop III.2$_{\rm D}$ cases (Fig. \ref{fig:plot_z-SFRD}a). 
In comparison to the background FUV field calculated by
\citet{agarwal12} and \citet{johnson13},
the local $J_{21}$ is at the same levels for $z \gtrsim 15$ but
starts to decline earlier for $z \lesssim 15$.
This earlier decline is because of our ignorance of Pop II stars 
for calculating the FUV radiation fields.

\section{Discussion}
\label{sec:discussion}

\subsection{The Final Fate of Primordial Stars}

The derived stellar mass distribution can be used to predict the fates of primordial stars.
\cite{yoon12} perform stellar evolution calculations and 
categorize the final fates of primordial stars as a function of the stellar masses. 
We classify our data of primordial stellar masses into five different populations 
depending on their final fates for non-rotating cases: 
neutron stars (NSs; $M_{*}/M_\odot<25$), 
low-mass black holes (LMBHs; $25<M_{*}/M_\odot<80$), 
pulsational pair-instability supernovae (PPISN; $80<M_{*}/M_\odot<120$), 
pair-instability supernovae (PISN; $120<M_{*}/M_\odot<240$), and 
high-mass black holes (HMBHs; $240<M_{*}/M_\odot$). 
Table \ref{tab:FinalFate} summarizes the respective number fractions and their redshift-dependence. 
Because the mass distribution itself evolves over time, 
the relative occurrence of the various fates of primordial stars also changes over cosmic time.


First of all, a large fraction of the primordial stars end their lives as HMBHs. 
They do not contribute to the early chemical evolution but leave remnant black holes 
that might have seeded the formation of supermassive black holes \citep{li07}. 
A fraction of Pop III.1$_{\rm H_2}$ stars die as PISN 
that chemically pollute the surrounding inter-stellar medium,
leaving the peculiar elemental abundance patterns associated with this type of supernovae. 
It has been a long-standing puzzle, however, that such distinct abundance patterns are not observed in Galactic metal-poor stars. 
Very recently, however, \cite{aoki14} report the discovery of a metal-poor star whose elemental abundance patterns can be explained 
by a supernova explosion of a very massive star with $M_* > 100 \ \msun$. 
Intriguingly, they also estimate from their data obtained from the sloan extension for galactic understanding and exploration (SEGUE) survey 
that such peculiar stars are rare, comprising only a few percent of metal-poor stars, 
in reasonable agreement with our simulation result.
One explanation for the challenges in finding PISN signatures may be 
that mini-haloes hosting such massive primordial stars have 
masses too low to retain the halo gas after an energetic supernova explosion \citep[e.g.,][]{cooke14}. 
Heavy elements produced in the stellar interior might be expelled to intergalactic space and are therefore absent from the gas 
out of which the metal-poor stars form. 
At lower redshifts, the Pop III.1$_{\rm HD}$ stars with $M_* \simeq 25 \ \msun$ die leaving LMBHs or NSs. 
Nucleosynthesis in core-collapse supernovae or hypernovae \citep[e.g.,][]{umeda03}, 
which are expected for such progenitors, are likely to dominate in the early Galactic chemical evolution.

\subsection{Uncertainty in Mass Estimation}
\label{ssec:uncertainties}


In the present investigation, we have used simple analytic formulae for estimating the stellar mass from local properties 
such as the gas infall rate $\dot{M}_{\rm Jeans}$ and the FUV intensity $J_{21}$. 
It would be important to improve the estimates by including physical effects that have not been considered here.


For instance, recent 3D numerical simulations show that a circumstellar disc 
becomes gravitationally unstable and fragments \citep[e.g.,][]{clark11b,greif11,greif12,stacy13b}. 
This could lead to the formation of multiple stellar systems rather than a single star in each mini-halo. 
In this case, the available gas in the envelope is divided among multiple protostars, 
so that the stars in such a multiple system could have relatively lower masses than 
in the case of a single star \citep[e.g.,][]{peters10, susa14}. 
However, there is an opposite effect; a large fraction of the protostars could rapidly migrate inward 
due to gravitational torque, resulting in frequent stellar mergers at the cloud centre \citep[e.g.,][]{greif12,vorobyov13}. 
This more efficient accretion could enhance the formation of massive stars. 
The final stellar masses are determined by the balance of the above competing effects,
which would also be affected by stellar radiative feedback \citep[e.g.,][]{stacy12,susa13}. 
Moreover, recent magnetohydrodynamic simulations show that even a small amount of turbulence can amplify 
magnetic fields by a dynamo mechanism \citep{sur10,federrath11,sur12,turk12}. 
The existence of magnetic fields would increase angular momentum transfer in the disc via magnetic braking, preventing disc fragmentation 
as well as enhancing accretion rates on to the protostars \citep[e.g.,][]{machida13}. 
Simulating the long-term evolution including all of these effects is still challenging, but should be tackled in future studies.

\begin{figure}
\begin{center}
\resizebox{8.5cm}{!}{\begin{tabular}{cc}
\includegraphics[clip,scale=1]{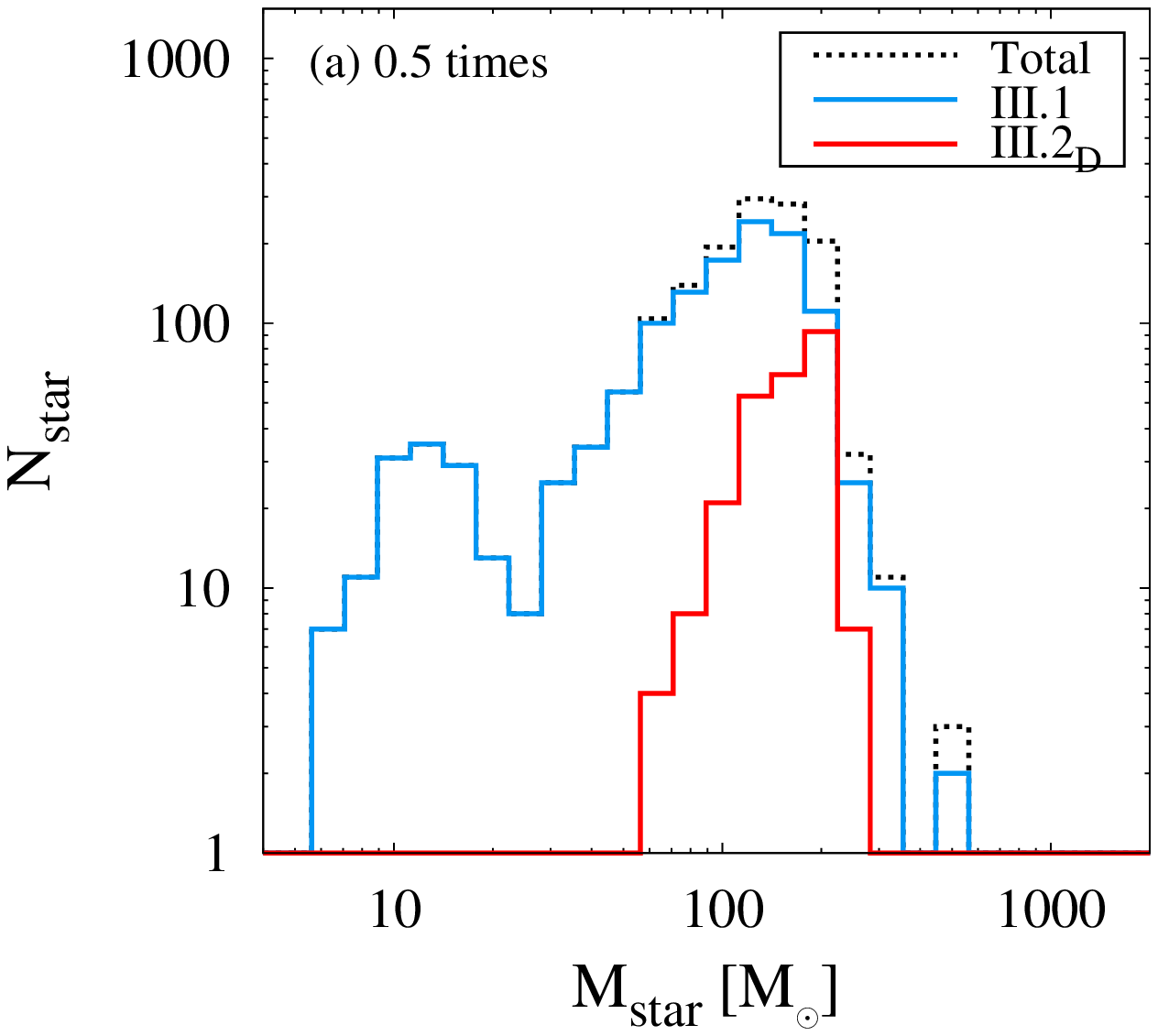}&
\hspace{-15mm}
\includegraphics[clip,scale=1]{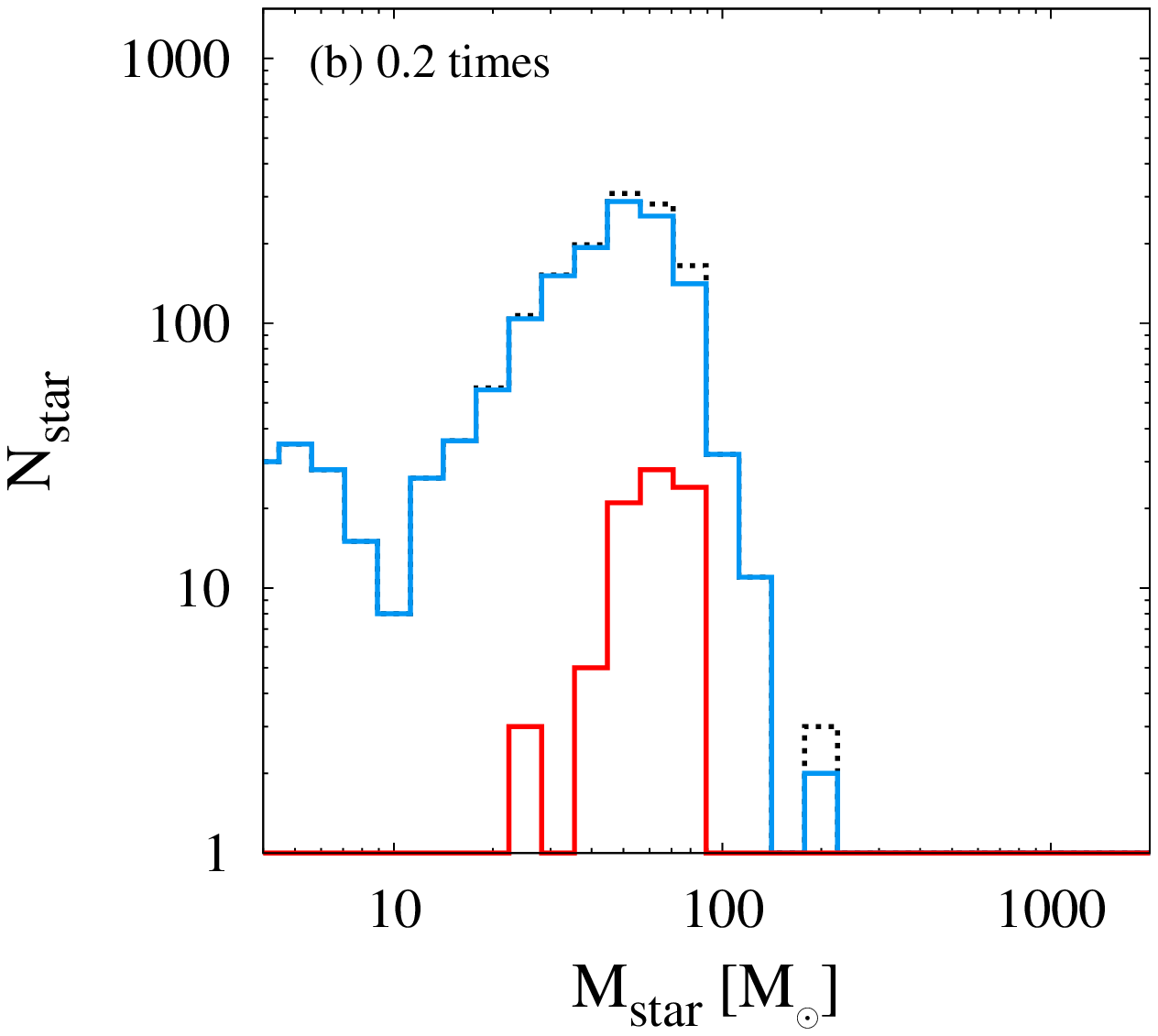}
\end{tabular}}
\caption{
Same as Fig. \ref{fig:plot_IMF} 
but calculated by decreasing the estimated stellar masses obtained from Eq. (\ref{eq:MIII_dMdt-Jeans})
by factors of 0.5 (panel a) and 0.2 (b).}
\label{fig:plot_IMF_compare}
\end{center}
\end{figure}


The number fraction of Pop III.2$_{\rm D}$ stars is dependent on 
the estimated mass of the Pop III.1 stars, 
which in turn determines the strength of the FUV radiation field. 
It is possible to see how critical this effect is by artificially modifying the scaling relation 
Eq. (\ref{eq:MIII_dMdt-Jeans}). Fig. \ref{fig:plot_IMF_compare} 
shows the mass distributions obtained using modified scaling relations, whereby
the masses obtained from Eq. (\ref{eq:MIII_dMdt-Jeans}) are multiplied by either 0.5 or 0.2. 
The local FUV intensity decreases because of the 
lower stellar masses, and the fraction of the clouds with $J_{21} > 0.1$ 
that form Pop III.2$_{\rm D}$ stars also decreases. 
In panel (a), the numbers of the two populations are still comparable 
at the high-mass end, whereas in panel (b), the overall mass distribution 
is dominated by Pop III.1 stars. 
Uncertainties in the masses of the Pop III.1 stars
affect the masses of the Pop III.2$_{\rm D}$ stars in this manner.


There are also uncertainties regarding the local FUV intensity. In this paper, 
we only have counted photodissociating radiation emitted by primordial stars. 
However, there could be other FUV emitters such as Pop II stars. 
Pop II stars dominate the cosmic star formation rate at low redshifts.
As noted in Section \ref{sec:method}, however, including their FUV radiation 
should not largely modify the stellar mass distributions 
because a large fraction of the primordial stars form at sufficiently high redshifts.


There is another type of Pop III.2 star that is formed under the influence of external radiation, 
the so-called Pop III.2$_{\rm I}$ star (see also Sec. \ref{sec:introduction}). 
We have indeed found potential sites of Pop III.2$_{\rm I}$ star formation, 
i.e., stars forming in gas clouds that were previously ionized by radiation from nearby stars. 
In principle, one can simulate the evolution of such clouds and nearby $\HII$ regions 
by solving the transfer of EUV ($h\nu > 13.6~{\rm eV}$) radiation \citep[e.g.,][]{yoshida07}. 
This will likely add an additional population of lower mass primordial stars 
\citep[e.g., $M_* =17 \ \msun$;][]{hosokawa12b}, which can be verified in future studies.
However, for the stellar mass distribution at the redshifts of interest in this paper ($10 < z < 30$), 
Pop III.2$_{\rm I}$ stars are a minor component and do not affect the conclusions of this study.

\subsection{Cosmological Samples}


There is a variety of physical processes involved in the formation of primordial stars.
Most of the uncertainties we have discussed are attributed to the complexity of the star formation process itself, 
in particular, during the stage of mass accretion on to the embryo protostar and the protostar's UV feedback on the
collapsing parent cloud.
By contrast, the early evolutionary phases until the formation of a primordial gas cloud is relatively well understood \citep{yoshida08,greif12} and 
our results provide statistics of the physical properties of the cosmological primordial gas clouds, 
such as the radial profiles of density, temperature, and velocities. 
In future, we shall discuss the variety of cloud properties and resultant stellar mass. 
Furthermore, we find that there is 
a certain normal initial condition of the primordial star formation
and its redshift-dependence.


Previous studies adopted different initial conditions and used different numerical codes with different physical processes implemented. 
It is thus often difficult to identify the origin of differences between studies. 
Direct comparison would be easier if the same initial conditions are used. 
Our sample of haloes can provide a convenient test set as well as statistical properties of haloes under normal initial conditions.\footnote{One of the authors (SH) is willing to provide data for individual sample haloes, or even for all the haloes to the interested reader.}

Ultimately, we will be able to determine the final stellar masses more accurately,
when we conduct detailed 3D radiation magnetohydrodynamic simulations, including the additional physics described in Section \ref{ssec:uncertainties}. 
Alternatively, we can do the same more efficiently with knowledge of the distribution of initial conditions.
Instead of repeating the entire multi-step procedure performed in the present study, 
we only need to update the estimation formula of the final stellar masses such as Eq. (\ref{eq:MIII_dMdt-Jeans}) 
by simulating the evolution for several representative cases. 
In this manner, the primordial stellar mass distribution can be easily reconstructed 
from a very large cosmological simulation.

\section*{Acknowledgements}
We thank Masahiro M. Machida, Hajime Susa, Kenji Hasegawa, and Kohei Inayoshi for stimulating discussions. 
We also thank the anonymous referee, whose careful comments improved the clarity of this paper.
The numerical calculations were carried out on Cray XC30 and the general-purpose PC farm at Center for Computational Astrophysics, CfCA, of National Astronomical Observatory of Japan, T2K-Tsukuba System at Center for Computational Sciences, University of Tsukuba, and SR16000 at YITP in Kyoto University. 
This work was supported by Grant-in-Aid for JSPS Fellows (SH) and by the Grants-in-Aid for Basic Researches by the Ministry of Education, Science and Culture of Japan (25800102: TH, 25287050: NY). 
Portions of this research were conducted at the Jet Propulsion Laboratory, California Institute of Technology, operating under a contract with the National Aeronautics and Space Administration (NASA).

\bibliography{biblio}
\bibliographystyle{mn2e}

\appendix

\section{Analytic Model for the Accreting Supergiant Protostar}
\label{app:spgp}


The evolution of mass accreting protostars greatly depends on the accretion rates. 
\cite{hosokawa12a} show that, with high accretion rates of $\dot{M} \gtrsim 0.04~\msunyr$, 
the protostellar evolution qualitatively differs from that at lower rates. 
At these high accretion rates, the stellar radius monotonically increases with increasing
the stellar mass, exceeding $10^3$ R$_\odot$ for $M_* > 100 \ \msun$. 
The UV radiative feedback from such ``supergiant protostars'' is
weak, because the resulting effective temperature is less than $10^4$ K. 
In the current work, we frequently see this evolutionary stage in our 2D RHD 
simulations of Pop III.2$_{\rm D}$ cases, 
where the accretion rates are relatively high. 
However, it is sometimes difficult to construct stellar models 
of these strongly accreting supergiant protostars by numerically solving the stellar structure equations,
especially with highly variable mass accretion histories.


Instead, we shall consider an analytic model for the evolution of the supergiant protostar. 
It is known that, during this evolutionary stage, the evolution of the stellar radius is well described by equation (11) in \cite{hosokawa12a},
\begin{eqnarray}
R_* (M_*) \simeq 260~{\rm R}_{\odot}\  \left( \frac{M_*}{{\rm M}_\odot} \right)^{1/2} \ ,
\label{eq:P3_Radius}
\end{eqnarray}
a formula derived analytically assuming the Eddington luminosity and a stellar surface temperature $T_{\rm eff} = 5000$~K. 
As shown in fig. 5 of \cite{hosokawa12a}, however, the stellar radius obtained using a detailed stellar evolution code is slightly less than that predicted by Eq. (\ref{eq:P3_Radius}). 
Nevertheless, the functional dependence with an exponent 1/2 is well matched. 
We therefore retain the functional form of this description but multiply the resulting radius by 0.8 to better fit the numerical results in Paper I, which also used a detailed stellar evolution code. 
The stellar effective temperature is nearly constant at $T_{\rm eff} = 5000$ K during this phase (see fig. 12 in Paper I), and the stellar luminosity is calculated from 
\begin{eqnarray}
L_* (R_*, T_{\rm eff} = 5000 \ {\rm K}) = 4 \pi R_*^2 \sigma T_{\rm eff}^4 \ , 
\end{eqnarray}
where $\sigma$ is the Stefan-Boltzmann constant. 
As long as the mass accretion rate is above the critical value $0.04 \ \msunyr$, 
we calculate the stellar radius and luminosity using the above equations. 
Once the accretion rate falls below this value, we switch to another analytic model of the ``oscillating protostar'' used in Paper I.


After the mass accretion rate falls below $4 \times 10^{-3} \ \msunyr$, the star begins to contract on a Kelvin-Helmholtz (KH) time-scale 
\citep[e.g.,][]{omukai03a},
\begin{eqnarray}
t_{\rm KH} = \frac{G M_*^2}{R_* L_*} \ .
\label{eq:KH-time}
\end{eqnarray}
The changes of stellar radius and luminosity during a timestep d$t$ are written as 
\begin{eqnarray}
\label{eq:P3_contraction_KH}
R_*^{\,\rm new} &=& R_* + \frac{{\rm d}t}{t_{\rm KH}} \cdot \left[ R(M_*)_{\rm ZAMS} - R_* \right] \ , \\
L_*^{\rm new} &=& L_* + \frac{{\rm d}t}{t_{\rm KH}} \cdot \left[ L(M_*)_{\rm ZAMS} - L_* \right] \ ,
\end{eqnarray}
where the properties of zero-age main sequence (ZAMS) stars are 
\begin{eqnarray}
R(M_*)_{\rm ZAMS} &=& 3.109 \times 10^{-1} \left( \frac{M_*}{{\rm M}_\odot} \right)^{0.58} \ , \\
L(M_*)_{\rm ZAMS} &=& 3.939 \times 10^{3} \left( \frac{M_*}{{\rm M}_\odot} \right)^{1.30} \ .
\label{eq:ZAMS}
\end{eqnarray}
If Eq.~\ref{eq:P3_contraction_KH} yields a value
$R_*^{\,\rm new}(M_*) < R_{\rm ZAMS}$, we use another relationship to get the stellar properties at the next step 
\begin{eqnarray}
R_*^{\,\rm new} &=& \left[ R_* + \frac{{\rm d}t}{t_{\rm KH}} R(M_*)_{\rm ZAMS} \right] \left( 1 + \frac{{\rm d}t}{t_{\rm KH}} \right)^{-1} ,\\
L_*^{\rm new} &=& \left[ L_* + \frac{{\rm d}t}{t_{\rm KH}} L(M_*)_{\rm ZAMS} \right] \left( 1 + \frac{{\rm d}t}{t_{\rm KH}} \right)^{-1} .
\label{eq:P3_contraction_analytic}
\end{eqnarray}

By using the above-mentioned formulae for protostars going through the
supergiant phase, we can compute the mass growth of a rapidly accreting 
Pop III.2$_{\rm D}$ star.

\label{lastpage}
\end{document}